\documentclass{preprint}

\usepackage{hyperref}       
\usepackage{url}            
\usepackage{booktabs}       
\usepackage{amsfonts}       
\usepackage{nicefrac}       
\usepackage{microtype}      
\usepackage{xcolor}         
\usepackage{siunitx}
\usepackage{titletoc}
\usepackage{array}
\usepackage{colortbl}
\usepackage{arydshln}

\usepackage{amsmath}
\usepackage{xspace}
\usepackage{graphicx}
\usepackage{xcolor}
\usepackage{booktabs}
\usepackage{makecell} 

\usepackage{framed}
\usepackage{amssymb}
\usepackage{color}
\usepackage{soul}

\usepackage[most]{tcolorbox}

\usepackage{enumitem}
\usepackage{multirow}

\usepackage{algorithm}
\usepackage{algpseudocode}
\usepackage{pifont}

\usepackage{siunitx}
\definecolor{DarkGreen}{RGB}{30,130,30}
\newcommand{\cmark}{\textcolor{DarkGreen}{\ding{51}}}
\newcommand{\xmark}{\textcolor{red}{\ding{55}}}%

\newtcbox{\hlwhite}{on line, box align=base, colback=red!20,colframe=white,size=fbox,arc=2pt, before upper=\strut, top=-3pt, bottom=-4.5pt, left=-2pt, right=-2pt, boxrule=0pt}

\definecolor{evidenceblue}{RGB}{233,242,255}

\definecolor{tagblue}{RGB}{100,150,200}

\definecolor{FBest}{RGB}{4,130,53}

\newcommand{\modelname}{LiveTradeBench\xspace}

\title{LiveTradeBench: Seeking Real-World Alpha with Large Language Models}

\author
{Haofei Yu, Fenghai Li, Jiaxuan You\\
\vspace{1em} 
\normalfont{\small University of Illinois, Urbana-Champaign}\\
\vspace{0.8em}
\href{https://trade-bench.live}{\includegraphics[height=0.9em]{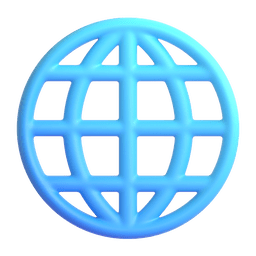}\xspace\texttt{trade-bench.live}}\\
\href{https://github.com/ulab-uiuc/live-trade-bench}{\includegraphics[height=0.9em]{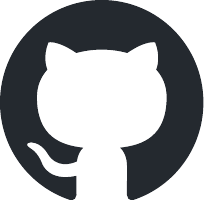}\xspace\texttt{github.com/ulab-uiuc/live-trade-bench}}
}

\begin{document}

\maketitle
\thispagestyle{firstpagestyle} 

\begin{abstract}
Large language models (LLMs) achieve strong performance across benchmarks—from knowledge quizzes and math reasoning to web-agent tasks—but these tests occur in static settings, lacking real dynamics and uncertainty. Consequently, they evaluate isolated reasoning or problem solving rather than decision-making under uncertainty. To address this, we introduce \modelname, a live trading environment for evaluating LLM agents in realistic and evolving markets. \modelname follows three design principles:
(i) Live data streaming of market prices and news, eliminating dependence on offline backtesting and preventing information leakage while capturing real-time uncertainty;
(ii) a portfolio-management abstraction that extends control from single-asset actions to multi-asset allocation, integrating risk management and cross-asset reasoning; and
(iii) multi-market evaluation across structurally distinct environments—U.S. stocks and Polymarket prediction markets—differing in volatility, liquidity, and information flow.
At each step, an agent observes prices, news, and its portfolio, then outputs percentage allocations that balance risk and return. Using \modelname, we run 50-day live evaluations of 21 LLMs across families. Results show that (1) high LMArena scores do not imply superior trading outcomes; (2) models display distinct portfolio styles reflecting risk appetite and reasoning dynamics; and (3) some LLMs effectively leverage live signals to adapt decisions. These findings expose a gap between static evaluation and real-world competence, motivating benchmarks that test sequential decision making and consistency under live uncertainty.
\end{abstract}

\section{Introduction}

Large language models (LLMs) have achieved near-saturation performance on diverse benchmarks--such as knowledge quizzes~\citep{hendrycks2020measuring,phan2025humanity,rein2024gpqa}, math reasoning tests~\citep{cobbe2021training,2023opencompass,quan2025codeelo}, and instruction-following tasks~\citep{pyatkin2025generalizing,zhou2023instruction,jiang2023followbench}.
However, these benchmarks are static, evaluating models on fixed inputs with single-turn reasoning. High scores on such tests do not necessarily reflect real-world intelligence, where agents must perceive, act, and adapt through feedback over time.

To move beyond static evaluation, recent work has introduced interactive environments that allow LLMs to perform sequential actions and observe feedback~\citep{jimenez2023swe,zhou2023sotopia}.
Examples include web and computer-use agents~\citep{zhou2023webarena,koh2024visualwebarena,xie2024osworld,he2024webvoyager}, which operate in discrete and deterministic environments—each action produces a predictable transition defined by backend logic.
These environments test perception and reasoning, but remain fully controllable and support tree-based searching~\citep{Koh2024TreeSFA,Putta2024AgentQAA,Aksitov2023ReSTMRA}. In contrast, trading environments represent continuous and autonomous systems.
The world evolves independently of the agent, and actions only adjust the agent’s internal portfolio state rather than directly determining future observations.
Feedback is delayed and noisy, emphasizing adaptation over control.
This difference in environment structure—from deterministic systems to dynamic processes—defines a deeper frontier for evaluating LLM agents’ ability to reason and act in open-ended, real-world settings~\citep{GarridoMerchan2024DeepRLA,Li2024INVESTORBENCHABA}.

\begin{table}[t!]
\centering
\caption{\textbf{Comparison of \modelname with existing trading benchmarks}. We compare our work with others through four dimensions: (1) \textit{sequential decision} for whether its current trading actions rely on the previous actions; (2) \textit{portfolio management} for whether its task is multi-asset portfolio management; (3) \textit{live trading} for whether the evaluation belongs to backtest with historical market data or live test with real-time streaming data; (4) \textit{multi-market evaluation} for whether it includes markets beyond the stock market.}
\begin{tabular}{lcccc}
\toprule
\textbf{Benchmark} &
\makecell{\textbf{Sequential}\\\textbf{Decision}} &
\makecell{\textbf{Portfolio}\\\textbf{Management}} &
\makecell{\textbf{Live}\\\textbf{Trading}} &
\makecell{\textbf{Multi-market}\\\textbf{Evaluation}} \\
\midrule
FinQA~\citep{chen2021finqa} & \xmark & \xmark & \xmark & \xmark \\
ConvFinQA~\citep{chen2022convfinqa} & \xmark & \xmark & \xmark & \xmark \\
FLUE~\citep{shah2022flue} & \xmark & \xmark & \xmark & \xmark \\
FinEval~\citep{zhang2023fineval} & \xmark & \xmark & \xmark & \xmark \\
BizFinBench~\citep{bigeard2025finance} & \xmark & \xmark & \xmark & \xmark \\
FinAgentBench~\citep{bigeard2025finance} & \xmark & \xmark & \xmark & \xmark \\
FinSearchComp~\citep{hu2025finsearchcomp} & \cmark & \xmark & \xmark & \xmark \\
INVESTORBENCH~\citep{Li2024INVESTORBENCHABA} & \cmark & \xmark & \xmark & \xmark \\
StockBench~\citep{chen2025stockbench} & \cmark & \cmark & \xmark & \xmark \\
DeepFund~\citep{Li2025TimeTIA} & \cmark & \cmark & \cmark & \xmark \\
\midrule
\textbf{LiveTradeBench (ours)} & \cmark & \cmark & \cmark & \cmark \\
\bottomrule
\end{tabular}
\end{table}

Despite this importance, current applications for building LLM-based trading agents remain oversimplified and disconnected from live market dynamics. Specifically, (1) most evaluation frameworks rely on offline backtesting, which is prone to information leakage and fails to capture the uncertainty, volatility, and feedback of real-world environments~\citep{Li2025TimeTIA,Li2025CanLFA,Li2025WillLBA,Papadakis2025StockSimADA}; and (2) most trading agents model trading as low-level local actions (\textit{e.g.}, buy/sell/hold) on a single asset, neglecting higher-level reasoning and planning across multiple assets~\citep{Ma2025AgentTAA,Briola2021DeepRLA,Han2023MasteringPTA,Han2023SelectATA}. This naturally raises a broader question: \textit{How can we effectively evaluate the trading ability of LLM-based agents under realistic market conditions at low cost?}

To answer this question, we introduce \modelname, a live trading environment designed to address both limitations above.
(1) \modelname streams live market data, financial news, and social signals, eliminating the dependence on offline backtesting and thereby avoiding information leakage from the root while capturing real-world uncertainty and feedback.
(2) It adopts the portfolio management abstraction, framing trading as a strategic allocation process that integrates risk management, temporal reasoning, and decision consistency across multiple assets~\citep{Yu2024FinConASA,Kou2024AutomateSFA,Gu2024AdaptiveAEA}.
At each step, the environment exposes dynamic observations—market conditions, contextual signals, and the agent’s historical decisions—and the LLM must output an updated portfolio allocation that balances risk and return over time.
By combining live data streaming with portfolio-level reasoning, \modelname offers a realistic, end-to-end platform to evaluate the true trading competence of LLM-based agents under evolving market dynamics.

Using this benchmark, we conduct two types of live trading evaluations: stock market (U.S. stocks) trading and prediction market (Polymarket\footnote{\url{https://polymarket.com/}}) betting.
We compare 21 mainstream LLMs across multiple model families and capability tiers.
Our analysis yields three key findings:
(1) State-of-the-art models in LMArena~\citep{chiang2024chatbot} do not exhibit state-of-the-art trading performance—high benchmark scores in general reasoning do not translate to superior trading outcomes;
(2) LLMs display distinct portfolio management styles, differing in their risk appetite, asset selection patterns, and allocation dynamics; and
(3) LLMs can effectively leverage real-time market and news signals to make more informed and adaptive trading decisions.
Together, these results reveal a disconnect between conventional LLM evaluation and real-world financial competence, motivating the development of more adaptive and robust portfolio management agents.

\section{Related Work}

\paragraph{Evaluation of trading agents}
The evaluation of LLM-based trading agents generally relies on three types of environments or benchmarks.
(1) Backtesting with historical market data is the mainstream approach~\citep{Xiao2024TradingAgentsMLA,Li2025CanLFA,Tian2025TradingGroupAMA,Tang2025AlphaAgentLAA}. However, such evaluations often suffer from information leakage~\citep{Li2025WillLBA,Li2025ProfitMRA} and poor generalization across longer or multi-regime market periods~\citep{gao2024finbench,jiang2025finagent}. To address these issues, several studies propose data contamination audits, entity anonymization, and temporal de-biasing protocols for fairer backtesting evaluation~\citep{he2024finmem,wu2025finllm}.
(2) Market simulators provide an alternative by constructing synthetic or self-designed trading environments~\citep{Emmanoulopoulos2025ToTOA,Zhang2024WhenAMA,Papadakis2025StockSimADA,Chen2023PutYMA,LopezLira2025CanLLA}. Yet, these simulators serve mainly as testbeds for behavioral analysis rather than producing realistic trading actions aligned with actual market dynamics.
(3) Live evaluation with real-time data represents an emerging direction. While widely explored in other domains such as question answering~\citep{Kasai2022RealTimeQWA,nie2025uqassessinglanguagemodels} and coding~\citep{liang2024livecodebench}, this approach remains largely unexplored in trading~\citep{Li2025TimeTIA}. Our work focuses on this live evaluation paradigm, which we argue offers the most faithful and future-proof assessment of LLM trading intelligence.

\paragraph{Action space design for trading agents}
The design of trading tasks varies substantially across objectives, which can be formalized through differences in the \textit{action space} of trading agents.
In stock markets, most LLM-based systems adopt a \textit{single-asset trading} formulation, where actions are discrete decisions such as buy, hold, or sell~\citep{Zhang2024AMFA,Li2023TradingGPTMSA,Gao2024SimulatingFMA,Ma2025AgentTAA,Zhang2025FinWorldAAA}.
While intuitive, this setup overlooks cross-asset dependencies and realistic portfolio interactions.
Other approaches focus on \textit{alpha prediction}, producing continuous vectors of alpha signals that represent expected excess returns or relative performance across assets~\citep{Islam2025TheEOA,Zhang2020AutoAlphaAEA,Sun2024CombiningTBA,Heinrich2021FactorIAA}.
However, these signals describe predictions rather than directly executable trading actions.
In betting markets, agents often output probability estimates for mutually exclusive outcomes (\textit{e.g.}, “Yes” vs. “No”)~\citep{Koning2022BettingMEA,DeHaven2024MinutebyMinuteFMA,Jumadinova2011AMPA}, which can be interpreted as implicit portfolio positions in complementary assets.
We unify these perspectives under a \textit{portfolio management} framework, where the agent outputs allocation ratios across multiple assets or outcomes~\citep{Ye2020ReinforcementLearningBPA,Sun2021ReinforcementLFA,Lucarelli2020ADQA}.
This formulation generalizes discrete trading, alpha prediction, and probabilistic betting within a single continuous decision space that naturally emphasizes risk–return trade-offs and inter-asset correlations.

\paragraph{Framework for LLM-based trading agents}
Various frameworks leverage LLMs to build trading agents in different styles. One line of research focuses on fine-tuning a single LLM with reinforcement learning (RL) to enhance decision-making and trading performance~\citep{wang2025tradingr1,Xiong2025FLAGTraderFLA,Koa2024LearningTGA,Zhang2025FinWorldAAA,Zha2025ANDA}. Another line explores multi-agent systems, where agents collaborate or compete through role differentiation to simulate realistic market dynamics~\citep{xu2024tradingagent,li2025contesttrade,zhang2025tradinggroup}.
In addition, capabilities such as tool use (e.g., API calls, data collectors)~\citep{Papadakis2025StockSimADA,Islam2025TheEOA}, self-reflection~\citep{Koa2024LearningTGA}, and memory~\citep{Yu2023FinMemAPA,Li2023TradingGPTMSA,Li2024INVESTORBENCHABA} have been recognized as key components for improving trading intelligence.
To provide a controlled yet extensible setup, we adopt a React-style~\citep{yao2022react} framework equipped with tool use and memory as our agent configuration.

\section{Building Live Trading Environment for Portfolio Management}
\label{sec:live-env}

\subsection{Definition of Portfolio Management}

\paragraph{Problem definition}
We formulate the portfolio management task as a partially observable Markov decision process (POMDP)
$\mathcal{E} = \langle \mathcal{S}, \mathcal{A}, \mathcal{O}, \mathcal{T}, \Omega \rangle$,
where $\mathcal{S}$ is the latent market state space, $\mathcal{A}$ the action space,
$\mathcal{O}$ the observation space, $\mathcal{T}$ the transition dynamics,
and $\Omega$ the observation emission function.
At each timestep $t$, the environment is in a latent state
$\mathbf{s}_t \in \mathcal{S}$, which encapsulates the true market condition,
including asset fundamentals, volatility, liquidity, and other unobserved factors.
The agent receives a partial observation
\begin{equation}
\mathbf{o}_t = (\mathbf{q}_t, \mathbf{p}_t, \mathbf{c}_t) = \Omega(\mathbf{s}_t),
\end{equation}
where $\mathbf{q}_t \in \mathbb{R}^{N}$ denotes the current asset holdings (including cash),
$\mathbf{p}_t \in \mathbb{R}^{N}$ the observable market prices,
and $\mathbf{c}_t$ contextual signals such as news, sentiment, or macro indicators.
The total portfolio value is computed as $v_t = \mathbf{q}_t^{\top}\mathbf{p}_t$.
Conditioned on the observation history $\mathbf{o}_{\le t}$,
the agent produces an action $\mathbf{a}_t \in \mathcal{A}$ representing
a target allocation vector, subject to $\sum_i a_t^{(i)} = 1$.

\paragraph{State transition function}
The environment transition captures the joint evolution of the market and the agent’s portfolio.It consists of two coupled processes:
an \textit{exogenous} market-state evolution, governed by real-world dynamics and observable as $(\mathbf{p}_t, \mathbf{c}_t) \rightarrow (\mathbf{p}_{t+1}, \mathbf{c}_{t+1})$, and an \textit{endogenous} portfolio adjustment induced by the agent’s allocation decision $\mathbf{a}_t$ under the new market state, leading to
$\mathbf{q}_t \rightarrow \mathbf{q}_{t+1}$.
Concretely, after executing $\mathbf{a}_t$, the market evolves according to $\mathcal{T}$,
producing new prices $\mathbf{p}_{t+1}$ and contextual signals $\mathbf{c}_{t+1}$.
The portfolio is revalued and rebalanced under the new prices as
\begin{equation}
v_{t+1}^- = \mathbf{q}_t^{\top} \mathbf{p}_{t+1}, \qquad
\mathbf{q}_{t+1} = v_{t+1}^- \frac{\mathbf{a}_t}{\mathbf{p}_{t+1}}, \qquad v_{t+1} = \mathbf{q}_{t+1}^{\top} \mathbf{p}_{t+1} = v_{t+1}^-
\label{transition}
\end{equation}
where division is element-wise.
The next observation $\mathbf{o}_{t+1}$ is emitted by $\Omega$,
capturing the updated portfolio state and market context. Since we focus on highly liquid assets such as Nvidia (NVDA) stocks and trending Polymarket markets to make up the portfolio, the agent’s individual trades exert negligible influence on real-world prices.
This assumption justifies modeling the simulated transition function $\mathcal{T}$ as closely aligned with the real world.

\paragraph{From allocation decisions to executable trading actions}
At each timestep $t$, after outputing the allocation decision $\mathbf{a}_t$, the agent can update its holdings from $\mathbf{q}_t$ to $\mathbf{q}_{t+1}$ through executable trading actions (BUY/SELL/HOLD). The executed trade vector is defined as $\Delta \mathbf{q}_t = \mathbf{q}_{t+1} - \mathbf{q}_t$, where a positive $\Delta q_t^{(i)} > 0$ indicates buying $\Delta q_t^{(i)}$ shares of asset $i$, a negative $\Delta q_t^{(i)} < 0$ indicates selling $|\Delta q_t^{(i)}|$ shares, and $\Delta q_t^{(i)} = 0$ corresponds to holding the current position. Once these trades are executed, the portfolio transitions to the new holdings \(\mathbf{q}_{t+1}\), and the total portfolio value \(v_{t+1}\) is updated according to Eq.~\ref{transition}.
This formulation provides a direct mapping from the high-level allocation action space to explicit buy, sell, and hold operations, without modeling low-level order execution mechanics.

\begin{figure}
    \centering
    \includegraphics[width=\linewidth]{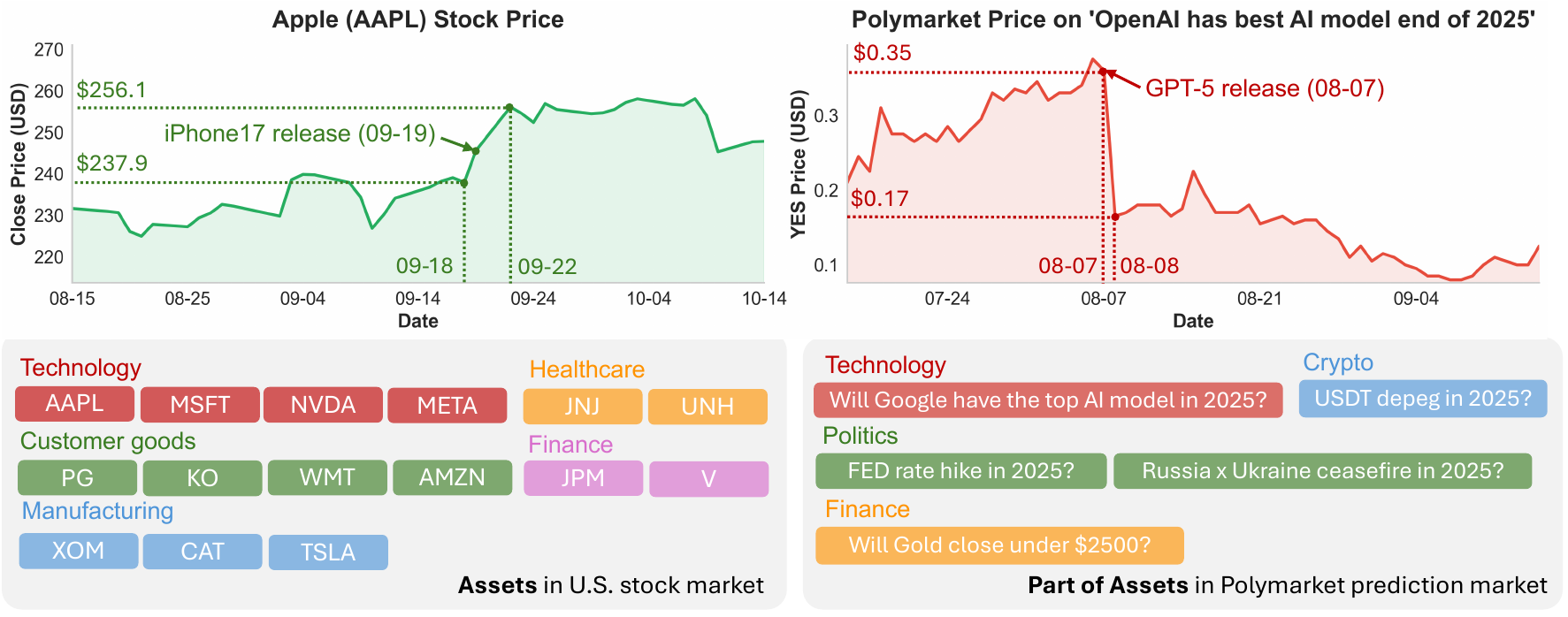}
    \caption{\textbf{Market selection in \modelname.} The top panels show AAPL in the U.S. stock market (left) and the contract “OpenAI has the best AI model by the end of 2025” in the Polymarket prediction market (right). In prediction markets, the price directly reflects the probability of a given outcome. Both markets respond to news and historical price trends, but Polymarket exhibits sharper fluctuations, faster reactions, and higher sensitivity to external signals. The bottom panels display representative assets across various domains, including technology, finance, cryptocurrency, manufacturing, and politics.}
    \label{fig:trade-bench-market}
\end{figure}

\subsection{Market Selection}
We evaluate agents in two complementary environments (stock market and prediction market) designed to test their generalization across both structured and information-driven regimes.  
This dual setup enables a comprehensive assessment of whether agents can perform consistently across markets that differ in structure, information flow, and reaction speed.  
Importantly, these two markets demand distinct strategies and reasoning perspectives for profitability:  
the stock market rewards long-horizon analysis and disciplined diversification,  
whereas the prediction market requires short-horizon adaptation and event-driven belief updating.

\paragraph{U.S. stock market}  
The U.S. stock market represents a mature, institutionally regulated system where asset prices evolve smoothly, exhibit strong cross-sector correlations, and reflect aggregated fundamentals and macroeconomic signals over time.  
Effective portfolio management in this environment requires capturing long-term dependencies, modeling hidden correlations, and maintaining diversified risk exposure.  
We construct a representative portfolio of 15 equities and ETFs spanning major U.S. sectors to ensure diverse responses to external information and macroeconomic shifts.  
The portfolio includes technology stocks—Apple (AAPL), Microsoft (MSFT), NVIDIA (NVDA), and Meta Platforms (META); financial stocks—JPMorgan Chase (JPM) and Visa (V); energy and industrial stocks—Exxon Mobil (XOM), Caterpillar (CAT), and Tesla (TSLA); consumer goods stocks—Procter \& Gamble (PG), Coca-Cola (KO), Amazon (AMZN), and Walmart (WMT); and healthcare stocks—Johnson \& Johnson (JNJ) and UnitedHealth Group (UNH).  
In addition, a \emph{cash asset} with a constant unit price and zero return rate is included to represent risk-free capital allocation.  
This composition provides balanced exposure across key sectors in a highly liquid and regulated financial environment, and we collect real-time stock prices as the data source for evaluation.

\paragraph{Polymarket prediction market}  
In contrast, the Polymarket prediction market is decentralized, sentiment-driven, and characterized by loosely coupled contracts that respond sharply and asynchronously to real-time information.  
These markets often move more abruptly and less coherently than stocks, reflecting shifts in collective belief rather than fundamentals.  
As a result, effective portfolio management here demands rapid adaptation, event-driven reasoning, and sensitivity to evolving narratives.  
We continuously track ten active binary prediction markets from \textit{Polymarket}, focusing on betting markets related to politics, crypto, technology and finance—such as “Fed rate hike in 2025?”, “Tether insolvent in 2025?”, “U.S. recession in 2025?”, and “USDT depeg in 2025?”.  
We hypothesize that prediction markets and stock markets respond to the same information with different latency and magnitude: stock markets integrate signals gradually through institutional consensus, while prediction markets react instantly and often overshoot due to speculative sentiment.  
Together, the two environments—structured financial markets and decentralized prediction markets—offer complementary testbeds for evaluating agents under both stability and uncertainty.

\begin{figure}
    \centering
    \includegraphics[width=\linewidth]{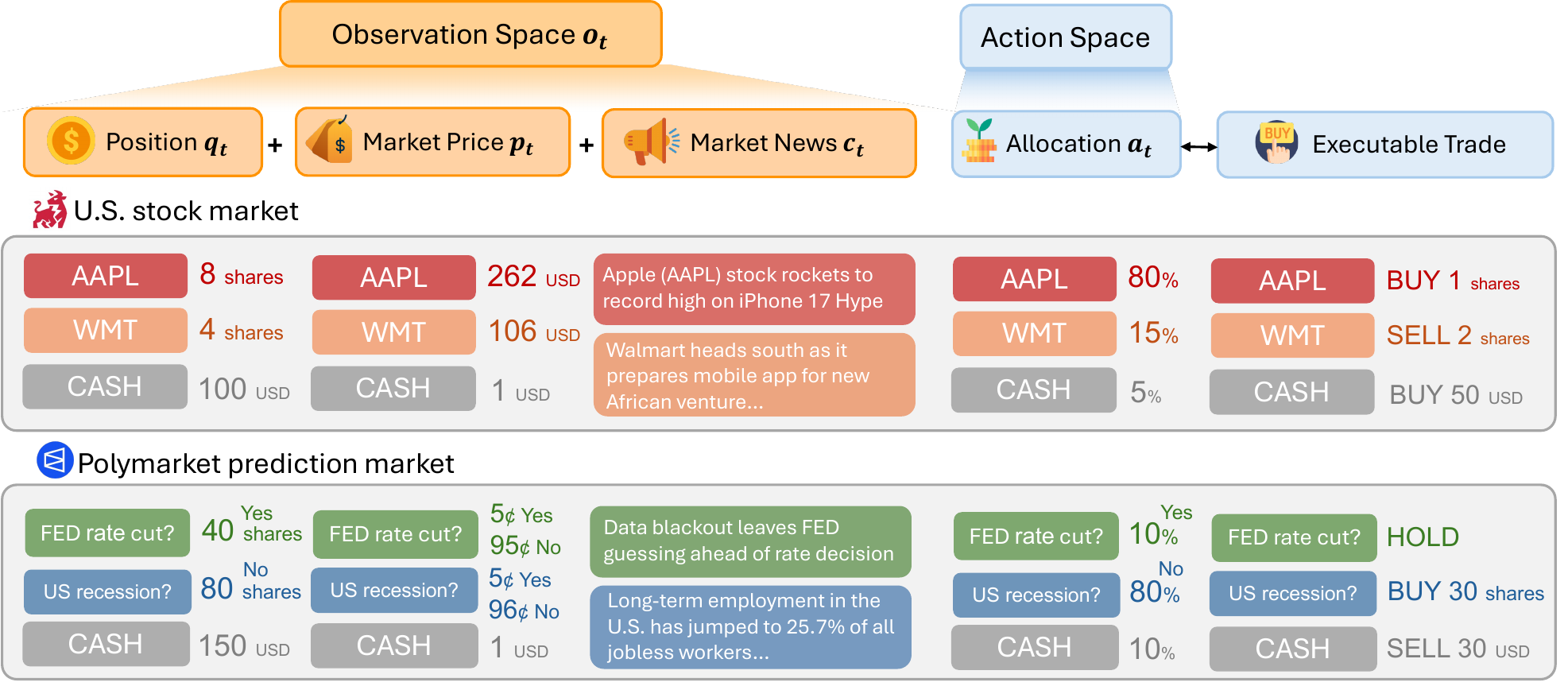}
    \caption{\textbf{Observation and action space for \modelname}. We illustrate examples from both the U.S. stock market and the Polymarket prediction market to demonstrate the observation and action spaces. The observation space consists of three components: the agent’s position, market prices, and relevant news context. The action space represents the portfolio allocation decisions generated by the agent, which can be directly translated into executable trading actions.}
    \label{fig:observation-and-action}
\end{figure}

\subsection{Observation Space}
At each timestep, the agent receives an observation $\mathbf{o}_t = (\mathbf{q}_t, \mathbf{p}_t, \mathbf{c}_t)$ that encapsulates the current market condition, external context, and portfolio status.  
This observation serves as the primary input for the agent’s decision-making process, integrating quantitative market signals, position information, and qualitative contextual cues.  
Based on these dynamic inputs, the agent determines its next allocation action, adapting to evolving market trends and external developments. Details on the data collection process of the observation space are available in Appendix~\S\ref{data-collection}.

\paragraph{Position $\mathbf{q}_t$}  
The position observation $\mathbf{q}_t$ represents the agent’s current holdings across all assets, including cash. Each component $q_t^{(i)}$ is a continuous, non-negative value indicating the number of units (or fraction thereof) of asset $i$ currently held in the portfolio. This formulation differs from discrete buy/hold/sell signals used in traditional trading formulations and instead provides a fine-grained representation of continuous capital allocation. The non-negativity constraint ensures that the agent cannot take short positions, reflecting realistic market restrictions and emphasizing capital distribution among long-only assets.

\paragraph{Market price $\mathbf{p}_t$}  
The price observation includes the latest asset prices and corresponding timestamps for all instruments in the portfolio. For the stock market, $\mathbf{p}_t$ contains closing prices of the 15 selected equities; for the prediction market, it includes the real-time trading prices of 10 trending \textit{Polymarket} markets. These values serve as the direct basis for portfolio valuation and allocation updates, enabling the agent to track how asset values evolve over time.

\paragraph{Market context $\mathbf{c}_t$}  
The contextual observation $\mathbf{c}_t$ mainly provides real-time market news. We collect recent articles from Google News using asset- and topic-specific keywords (e.g., “Federal Reserve,” “inflation,” “NVIDIA stock”). Such information reflects short-term market sentiment, investor attention, and macro-level signals that often precede price movements. To enable the model to reason about these factors, we include the textual summaries of these news items directly in the prompt, allowing LLM-based agents to incorporate qualitative context—such as sentiment shifts, policy expectations, or company-related events—into their trading decisions.

\subsection{Action Space}

At each timestep $t$, the agent makes an action $\mathbf{a}_t \in \mathcal{A}$, where $\mathcal{A}$ denotes the probability simplex action space. Each component $a_t^{(i)}$ specifies the proportion of the total portfolio value $v_t$ allocated to asset $i$, satisfying the budget constraint $\sum_i a_t^{(i)} = 1$. By default, we assume a long-only setting where $a_t^{(i)} \ge 0$ for all $i$. This continuous allocation-based formulation abstracts away low-level trading execution and focuses on high-level portfolio rebalancing, allowing agents to express smooth strategic shifts over time and directly optimize for portfolio-level objectives such as return, risk, and stability.

\paragraph{Stock market action}  
In the stock market environment, each action component $a_t^{(i)}$ represents the percentage of the total portfolio value $v_t$ to be allocated to stock $i$.  
This allocation determines the post-trade position $\mathbf{q}_{t+1}$ through proportional rebalancing and directly reflects the agent’s capital distribution across sectors.

\paragraph{Prediction market action}  
In the prediction market environment, each binary contract corresponds to two complementary assets—YES and NO. For $k$ active markets, the action vector $\mathbf{a}_t$ has $2k$ components, where $a_{t,\text{YES}}^{(k)}$ and $a_{t,\text{NO}}^{(k)}$ denote the portfolio allocations to the YES and NO outcomes of market $k$, respectively. The agent’s net exposure is defined as $e_t^{(k)} = a_{t,\text{YES}}^{(k)} - a_{t,\text{NO}}^{(k)}$, where a positive value indicates higher confidence in the YES outcome.

\section{Designing LLM-based Agents for Portfolio Management}
\label{sec:framework}

\begin{figure}
    \centering
    \includegraphics[width=\linewidth]{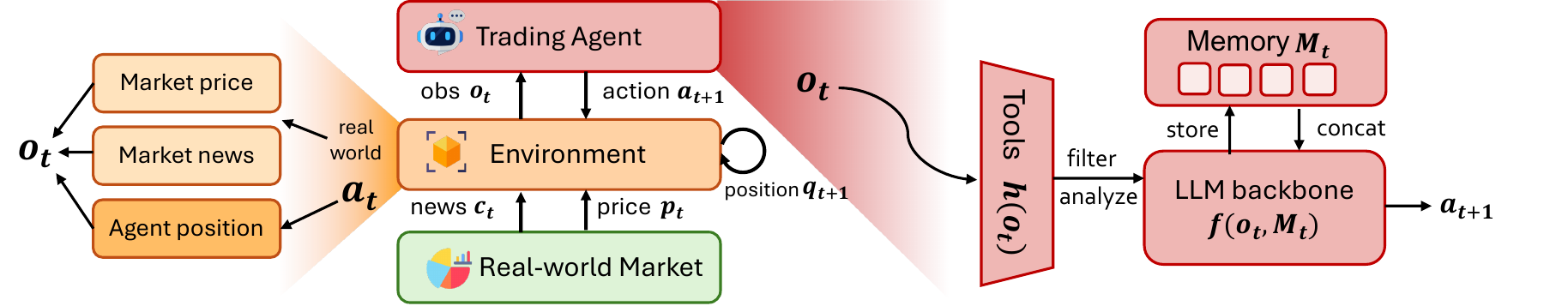}
    \caption{\textbf{Agent and environment framework in \modelname}.
The left side illustrates the simulated environment, which continuously retrieves real-world market prices and news, updating its internal state accordingly. It also adjusts the agent’s portfolio position based on the executed actions.
The right side depicts the portfolio-management agent, equipped with analytical tools to process observations from the environment. The agent maintains a memory of past observations, enabling adaptive and context-aware decision-making.}
    \label{fig:trade-bench-agent}
\end{figure}

In our framework, the \textbf{agent} is the central decision-making entity that transforms observed information into actionable portfolio allocations.
It serves as the bridge between the external market and its internal portfolio memory, continuously adapting its strategy to changing conditions.
The agent integrates three intertwined capabilities—\emph{tool use}, \emph{memory}, and \emph{reasoning}—that together enable it to perceive, recall, and act, forming a closed loop of information acquisition, reflection, and execution.
Formally, at each timestep $t$, the agent receives an observation $\mathbf{o}_t$.
In addition, the agent maintains an internal memory state $\mathbf{M}_t$, which stores the past observation beyond the current one.
Conditioned on both the current observation and memory, the agent produces an allocation
\begin{equation}
\mathbf{a}_t = f_\theta(\mathbf{o}_t, \mathbf{M}_t),
\end{equation}
where $f_\theta$ is a parameterized policy defining the agent’s trading behavior. Details on how we construct the prompt for the agent are available in Appendix~\S\ref{prompting-details}.

\vspace{1mm}
\paragraph{Tool use}
The first tool-use component enables the agent to interact with the live environment we provide—fetching, filtering, and extract real-time market and contextual information.  
While market prices $\mathbf{p}_t$ and contextual signals $\mathbf{c}_t$ are emitted by the environment, the tool-use module governs how the agent actively acquires and processes them.  
It acts as the agent interface with the real world, transforming raw inputs into structured feature representations
\begin{equation}
\mathbf{\tilde{o}}_t = h(\mathbf{o}_t),
\end{equation}
that capture both quantitative dynamics (e.g., price changes, returns, and volatility features derived from $\mathbf{p}_t$) and qualitative cues (e.g., news relevant to specific markets extracted from $\mathbf{c}_t$).  
Through tool use, the agent extends its perception beyond static observations, dynamically gathering and refining evidence to inform its allocation decisions.

\vspace{1mm}
\paragraph{Memory}
The second memory component maintains a compact representation of the agent’s recent observations and the outcome of its actions.  
At each timestep, the agent stores a fixed-length sequence of past observations and concatenates them into a unified memory state:
\begin{equation}
\mathbf{M}_t = \{\mathbf{o}_{\tau} \mid t - \Delta \le \tau < t\},
\end{equation}
where $\Delta$ denotes the memory horizon.  
This concatenated memory provides temporal context beyond the current observation, enabling the agent to capture dependencies such as volatility dynamics, allocation adjustments, and drawdown trends.  
By conditioning its decisions on both $\mathbf{o}_t$ and $\mathbf{M}_t$, the agent becomes adaptive to evolving market conditions over time.

\vspace{1mm}
\paragraph{Reasoning}
The third reasoning component serves as the agent’s decision core.  
It integrates information gathered through tool use with contextual knowledge retained in memory, forming a coherent understanding of the market at each moment.  
Before executing any action, the agent engages in a reasoning process that follows the ReAct~\citep{yao2022react} framework, where it first generates intermediate thoughts to interpret signals, recall relevant experiences, and hypothesize about potential outcomes.  
Similar to chain-of-thought prompting~\citep{wei2022chain}, this step produces explicit reasoning traces that connect perception and decision.  
Such interpretability allows the resulting actions to be analyzed and considered as rational responses to evolving market contexts.  
Through this deliberate reasoning–then–acting cycle, the agent achieves both adaptability and transparency in portfolio management. Such reasoning rationales can be potentially used to help researchers understand the model behavior and utilized as resources to improve LLMs.

\section{Evaluating LLM-based Agents under Live Test}
\label{sec:baselines}
In this section, we present detailed evaluation results and analyses of live trading conducted from \textbf{August 18, 2025}, to \textbf{October 24, 2025}—a total of 50 trading days—across trading agents built on 21 unique LLM backbones.
Section~\S\ref{backbone_llm} and ~\S\ref{trading_metric} describe the evaluation setup, including model backbones and performance metrics.
Section~\S\ref{eval_results} reports the main results, and Section~\S\ref{analysis_discussion} provides in-depth analyses and discussions.

\vspace{-1mm}
\subsection{Backbone LLMs for Evaluation}
\vspace{-1mm}
\label{backbone_llm}

To benchmark performance in the live trading environment, we evaluate a diverse set of mainstream LLMs as trading agents.
Specifically, we consider six representative model families.
These models are selected based on two main criteria:
(1) their state-of-the-art performance on general-purpose reasoning, knowledge and agentic benchmarks, and
(2) their diversity in model size, architecture, and performance levels, which allows us to study performance gradients across heterogeneous systems in financial decision-making tasks.

\paragraph{LLM family} We include the following representative models:
{Claude family} (Claude-Sonnet-3.7~\citep{anthropic2024claude3}, Claude-Opus-4 \& Claude-Sonnet-4~\citep{anthropic2024claude4}, Claude-Opus-4.1~\citep{anthropic2024claude41}),
{Grok family} (Grok-3~\citep{grok3}, Grok-4~\citep{grok4}),
{Qwen family} (Qwen2.5-72B-Instruct~\citep{Yang2024Qwen25TR}, Qwen3-235B-A22B-Instruct \& Qwen3-235B-A22B-Thinking~\citep{yang2025qwen3}),
{LLaMA family} (Llama3.3-70B-Instruct-Turbo~\citep{llama33}, Llama4-Scout \& Llama4-Maverick~\citep{llama4}),
{GPT family} (GPT-5~\citep{gpt5}, GPT-4o~\citep{hurst2024gpt}, GPT-4.1~\citep{gpt41}, GPT-o3~\citep{gpto3}),
{Kimi family} (Kimi-K2-Instruct~\citep{team2025kimi}), and
{DeepSeek} family (DeepSeek-V3.1~\citep{deepseek31}, DeepSeek-R1~\citep{guo2025deepseek}). Each model is wrapped in the same agentic framework that converts market observations into natural-language prompts and parses model outputs into structured portfolio allocation vectors.
This unified setup ensures that performance differences primarily reflect the models’ intrinsic reasoning and decision-making abilities rather than disparities in prompt formatting or execution. Details on the model selection are available in Appendix~\S\ref{model_details}.

\subsection{Trading Metrics for Evaluation}
\label{trading_metric}

To evaluate the performance of trading agents, we employ four widely used financial metrics: cumulative return, volatility, Sharpe ratio, and maximum drawdown (MDD).  
These metrics jointly capture profitability, risk exposure, risk-adjusted efficiency, and downside protection, offering a comprehensive view of trading performance across different markets.

\paragraph{Cumulative return ($CR = \frac{v_T - v_0}{v_0}$)}  
It measures the overall profitability of an investment strategy over a given evaluation period.  
Here, $v_0$ and $v_T$ denote the initial and final portfolio values, respectively, and $T$ is the total number of timesteps during evaluation.  
A higher cumulative return $CR$ indicates stronger cumulative gains achieved by the trading agent.

\paragraph{Sharpe ratio ($SR = \frac{\bar{r} - r_f}{\sigma}$)}  
It evaluates the efficiency of returns relative to the amount of risk taken.  
Here, $\bar{r}$ denotes the mean return, $r_f$ is the risk-free rate representing the baseline return from a no-risk investment, and $\sigma$ is the volatility of returns.  
In the U.S. stocks, $r_f$ corresponds to the short-term Treasury yield (typically positive), whereas in the Polymarket, $r_f$ is set to 0 to reflect the absence of yield on stablecoin-denominated assets.  
A higher Sharpe ratio $SR$ signifies that the strategy achieves greater reward per unit of risk, reflecting superior risk-adjusted performance.

\begin{table}[t]
\centering
\small
\caption{\textbf{Comparison of trading performance across U.S. stock and Polymarket prediction markets.} 
We use five key metrics: cumulative return (CR), Sharpe ratio (SR), maximum drawdown (MDD), win rate (WR), and volatility ($\sigma$). The highest value in each column is highlighted in bold.}
\begin{tabular}{l
S[table-format=1.2, table-column-width=0.8cm]
S[table-format=1.2, table-column-width=0.8cm]
S[table-format=1.2, table-column-width=0.8cm]
S[table-format=2.2, table-column-width=0.8cm]
S[table-format=1.2, table-column-width=0.8cm]
S[table-format=2.2, table-column-width=0.8cm]
S[table-format=1.2, table-column-width=0.8cm]
S[table-format=2.2, table-column-width=0.8cm]
S[table-format=2.2, table-column-width=0.8cm]
S[table-format=2.2, table-column-width=0.8cm]}
\toprule
\makecell[l]{\textbf{Model}} &
\multicolumn{5}{c}{\textbf{U.S. Stock Market}} &
\multicolumn{5}{c}{\textbf{Polymarket Prediction Market}} \\
\cmidrule(lr){2-6} \cmidrule(lr){7-11}
 & {CR}$\uparrow$ & {SR}$\uparrow$ & {MDD}$\downarrow$ & {WR}$\uparrow$ & {$\sigma$}$\downarrow$ &
   {CR}$\uparrow$ & {SR}$\uparrow$ & {MDD}$\downarrow$ & {WR}$\uparrow$ & {$\sigma$}$\downarrow$ \\
\midrule
Claude-Sonnet-3.7 & 3.63 & 1.45 & 2.65 & 59.18 & 10.25 & \textbf{20.54} & \textbf{2.38} & \textbf{10.65} & 51.02 & 44.64 \\
Claude-Sonnet-4 & 2.45 & 0.72 & 3.23 & 53.06 & 12.86 & -40.32 & -2.40 & 51.10 & 38.78 & 92.16 \\
Claude-Opus-4 & 3.93 & 1.72 & 2.11 & 63.27 & 9.45 & -2.04 & 0.09 & 13.67 & 46.94 & 56.38 \\
Claude-Opus-4.1 & 3.73 & 1.51 & \textbf{1.84} & 63.27 & 10.17 & -25.69 & -3.02 & 30.53 & 48.98 & 46.81 \\
\midrule
Grok-3 & 3.22 & 1.26 & 2.13 & 65.31 & 10.17 & -8.35 & -0.55 & 18.80 & 53.06 & 54.35 \\
Grok-4 & 4.30 & 1.75 & 1.92 & 59.18 & 10.43 & 7.38 & 1.01 & 13.04 & 46.94 & 46.92 \\
\midrule
Qwen2.5-72B-Instruct & 5.15 & 2.18 & 2.22 & 65.31 & 10.24 & 1.63 & 0.43 & \textbf{7.46} & \textbf{59.18} & \textbf{30.36} \\
Qwen3-235B-A22B-Instruct & 3.52 & 1.32 & 2.04 & 61.22 & 10.89 & -54.24 & -2.97 & 54.24 & 40.82 & 112.37 \\
Qwen3-235B-A22B-Thinking & 1.78 & 0.60 & 2.32 & 59.18 & \textbf{9.28} & -57.62 & -1.81 & 72.10 & 38.78 & 166.92 \\
\midrule
Llama3.3-70B-Instruct-Turbo & 2.72 & 0.88 & 3.54 & 61.22 & 11.95 & 1.58 & 0.40 & 19.41 & 57.14 & 38.15 \\
Llama4-Scout & 4.65 & 1.99 & 2.62 & 59.18 & 9.98 & -16.05 & -1.18 & 24.80 & 51.02 & 60.37 \\
Llama4-Maverick & 4.46 & 1.65 & 2.45 & 53.06 & 11.59 & -18.31 & -1.92 & 28.63 & 34.69 & 48.21 \\
\midrule
GPT-4o & 3.55 & 1.39 & 2.43 & 55.10 & 10.38 & -30.96 & -3.26 & 35.31 & 30.61 & 53.75 \\
GPT-4.1 & \textbf{6.25} & \textbf{2.64} & 1.92 & \textbf{65.31} & 10.51 & -33.69 & -1.74 & 37.98 & 40.82 & 95.27 \\
GPT-5 & 5.31 & 2.19 & 2.53 & 65.31 & 10.60 & -23.96 & -0.49 & 38.92 & 44.90 & 130.37 \\
GPT-o3 & 6.04 & 2.57 & 2.27 & 61.22 & 10.41 & -54.84 & -3.68 & 60.99 & 40.82 & 97.27 \\
\midrule
Gemini-2.5-Flash & 2.10 & 0.72 & 3.10 & 55.10 & 11.25 & -22.40 & -0.82 & 42.35 & 38.78 & 115.42 \\
Gemini-2.5-Pro & 1.95 & 0.61 & 2.85 & 50.00 & 10.98 & -35.15 & -1.65 & 49.80 & 34.69 & 101.87 \\
\midrule
Kimi-K2-Instruct & 3.07 & 1.15 & 3.32 & 53.06 & 10.53 & -53.44 & -5.26 & 54.74 & 28.57 & 69.41 \\ 
DeepSeek-V3.1 & 2.46 & 0.86 & 2.45 & 59.18 & 10.61 & -4.68 & -0.07 & 22.43 & 48.98 & 64.74 \\
DeepSeek-R1 & 2.10 & 0.78 & 2.20 & 61.22 & 9.11 & -13.19 & 0.14 & 44.16 & 42.86 & 143.25 \\
\bottomrule
\end{tabular}
\label{tab:market_comparison}
\end{table}

\paragraph{Maximum drawdown ($MDD = \max_{t \in [1, T]} \frac{\max_{i \in [1, t]} v_i - v_t}{\max_{i \in [1, t]} v_i}$)}  
It quantifies the largest observed decline from a historical peak to a subsequent trough in portfolio value before a new peak is reached.  
Here, $v_t$ represents the portfolio value at time step $t$.  
A smaller MDD indicates better downside protection and stronger resilience against severe losses.

\paragraph{Win rate ($WR = \frac{1}{T-1}\sum_{t=2}^{T}\mathbb{I}(r_t > 0)$)}
It measures the proportion of profitable trading steps, capturing the agent’s consistency in generating positive returns.
Here, $\mathbb{I}(\cdot)$ equals 1 when the return $r_t$ is positive and 0 otherwise.
A higher win rate $WR$ indicates that the agent achieves gains more frequently, complementing cumulative and risk-adjusted metrics by reflecting short-term decision reliability.

\paragraph{Volatility ($\sigma = \sqrt{\frac{1}{T-1}\sum_{t=1}^{T}(r_t - \bar{r})^2}$)}  
It reflects the variability of returns and serves as a measure of investment risk.  
Here, $r_t = \frac{v_t - v_{t-1}}{v_{t-1}}$ represents the return at time step $t$, $\bar{r} = \frac{1}{T}\sum_{t=1}^{T}r_t$ is the average return, and $T$ is the total number of evaluation timesteps.  
Strategies with lower volatility $\sigma$ exhibit more stable performance over time.

\subsection{Evaluation Results}
\label{eval_results}

\begin{figure}[t]
    \centering
    \begin{minipage}[t]{0.48\linewidth}
        \centering
        \includegraphics[width=\linewidth]{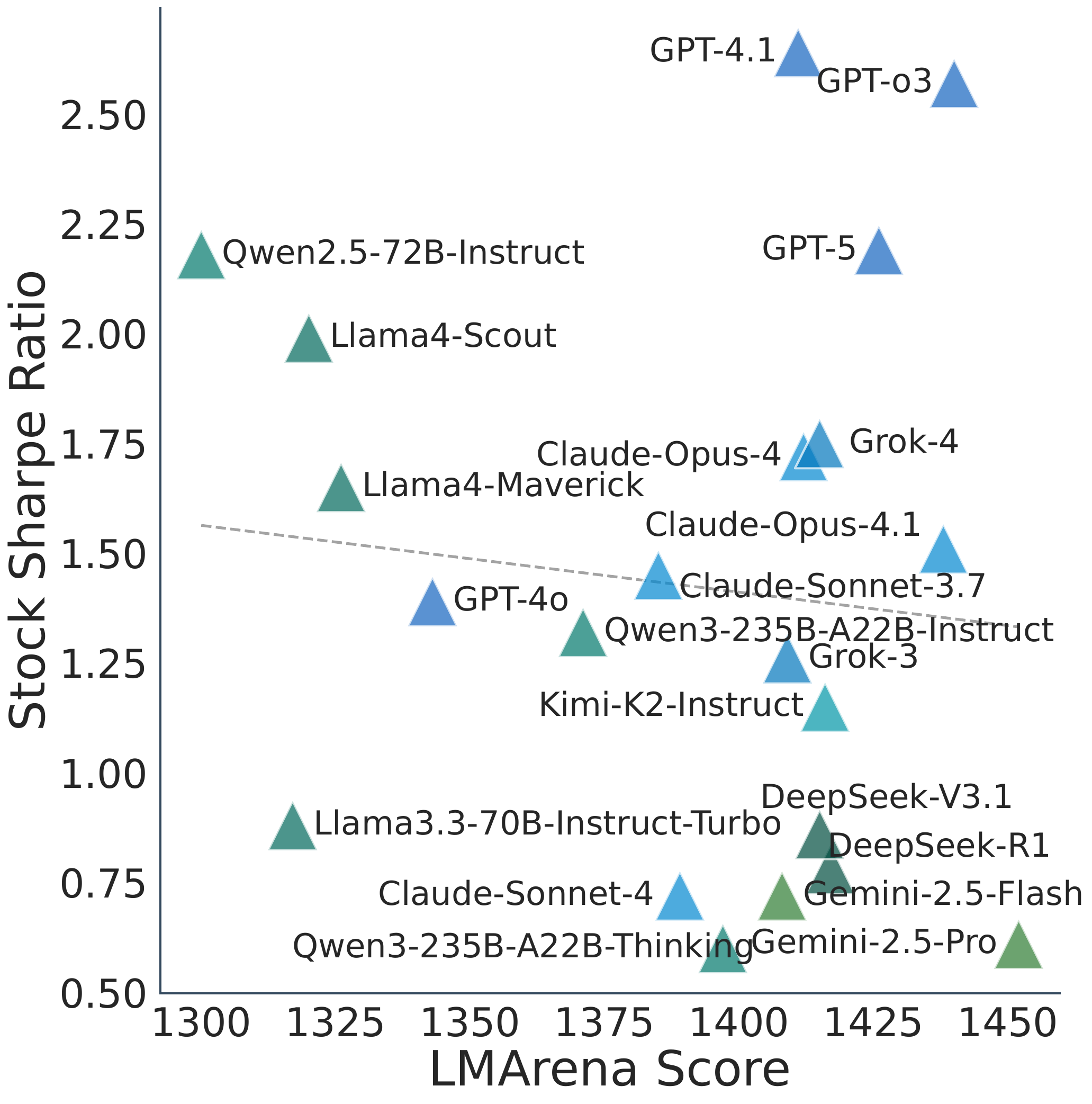}
    \end{minipage}
    \hfill
    \begin{minipage}[t]{0.48\linewidth}
        \centering
        \includegraphics[width=\linewidth]{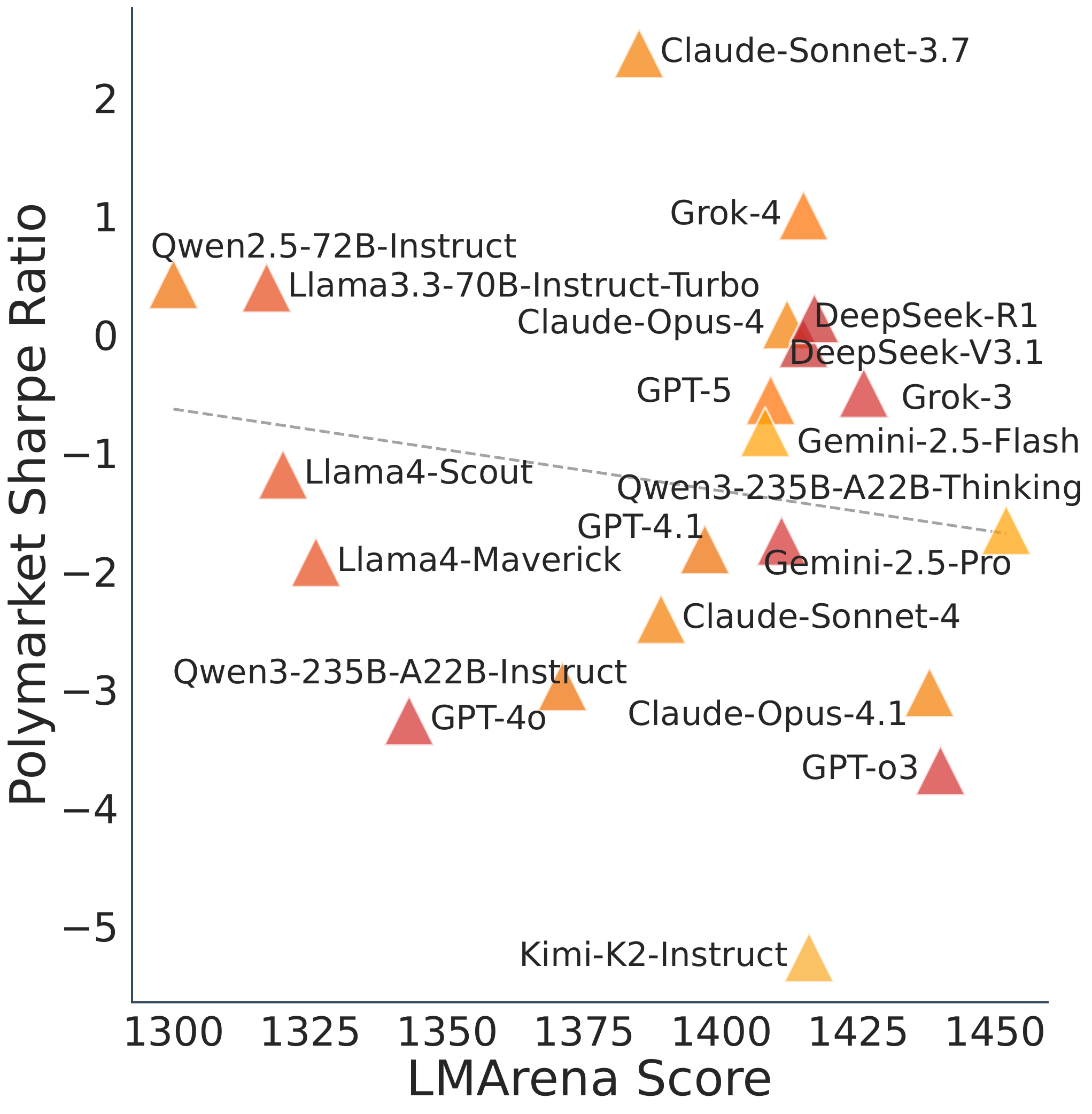}
    \end{minipage}
    \caption{\textbf{Correlation between LMArena score and Sharpe ratio across two markets.} 
    (\textbf{left}) U.S. stock market. (\textbf{right}) Polymarket prediction market. 
    Models from different families are shown in different colors, and the dashed line indicates the linear regression fit.}
    \label{fig:lmarena_correlation}
\end{figure}


\paragraph{Trading performance on one market does not generalize to another.}
As shown in Table~\ref{tab:market_comparison}, the Sharpe ratio correlation between the two markets is close to zero, indicating that success in one market does not imply success in the other. This highlights the need for market-specific trading strategies. For example, {Qwen2.5-72B-Instruct} and {Grok-4} show relatively consistent performance across both the stock and prediction markets, suggesting more stable and low-volatility strategies. In contrast, {GPT-4.1} achieves the highest cumulative return rate ($>6\%$) in the stock market but performs poorly in Polymarket (return $<-30\%$), likely due to overreactive allocation changes under higher volatility. Overall, the prediction market exhibits faster dynamics, greater volatility, and deeper drawdowns (MDD), demanding more agile and risk-tolerant strategies.

\paragraph{High general LLM capability does not imply strong financial performance.}
Figures~\ref{fig:lmarena_correlation} show that general LLM ability, as measured by LMArena scores, has slightly negative correlation with trading abilities. In the stock market, the Spearman correlation between LMArena scores and cumulative returns is only 0.054—virtually no relationship. In Polymarket, the correlation drops to –0.38, meaning models with higher general language ability often perform worse in the dynamic market. Thus, state-of-the-art LLMs on general benchmarks do not necessarily translate to state-of-the-art performance in dynamic, real-world trading. It highlights the uniqueness and necessity of our environment.

\paragraph{Distinct portfolio management styles emerge across models.}
Different models exhibit distinct management preferences. {Claude-Opus-4.1} and {Grok-4} adopt conservative strategies characterized by lower volatility and smaller drawdowns, prioritizing stability over aggressive gains. In contrast, {Kimi-K2-Instruct} and {GPT-5} display more risk-seeking behaviors—accepting higher volatility and MDD in pursuit of greater returns. Beyond return and risk metrics, models also differ notably in their portfolio composition and cash management patterns. For instance, {GPT-4o} consistently focuses on a few core assets (AAPL, MSFT, NVDA), whereas {GPT-5} diversifies across a broader range of stocks with smaller position ratios. Likewise, {Llama4-Scout} maintains a persistently high cash ratio (above 20\%), reflecting a more cautious liquidity stance, while {GPT-5} always keeps cash below 10\% throughout trading except in extremely high risk. These behavioral patterns are not limited to a single market—similar management styles emerge consistently across both markets.

\paragraph{Large reasoning models do not confer trading advantages.}
Consistent with findings from \citet{chen2025stockbench}, models explicitly designed for reasoning—such as {DeepSeek-R1}, {Qwen3-235B-A22B-Thinking}, and {GPT-o3}—do not outperform others in trading performance. Instead, they exhibit substantially higher volatility ($>$140 in Polymarket), implying over-adjustment during the decision process. The type of reasoning beneficial for mathematical or coding tasks does not straightforwardly transfer to financial or social reasoning. In fact, excessive deliberation observed in these models during trading can introduce instability and degrade trading consistency.

\subsection{Analysis and Discussion}
\label{analysis_discussion}

\begin{figure}[t]
    \centering
    \begin{minipage}[t]{0.32\linewidth}
        \centering
        \includegraphics[width=\linewidth]{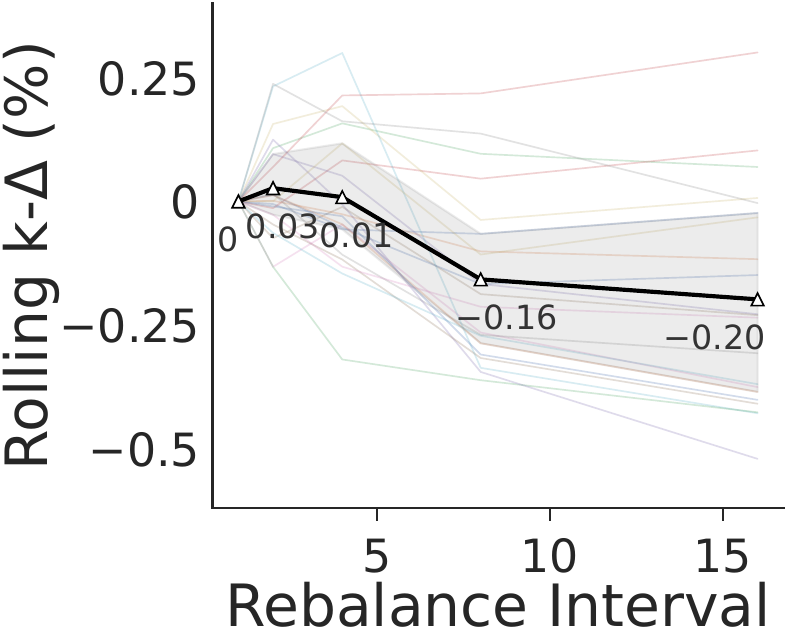}
        \caption{\textbf{Rolling $k$-delta analysis on U.S. stocks.} We evaluate rebalance intervals $k \in \{1, 2, 4, 8, 16\}$. The black line denotes the mean performance across 21 models, and the shaded gray region indicates the 25–75\% confidence interval.}
        \label{fig:delta_stock}
    \end{minipage}
    \hfill
    \begin{minipage}[t]{0.32\linewidth}
        \centering
        \includegraphics[width=\linewidth]{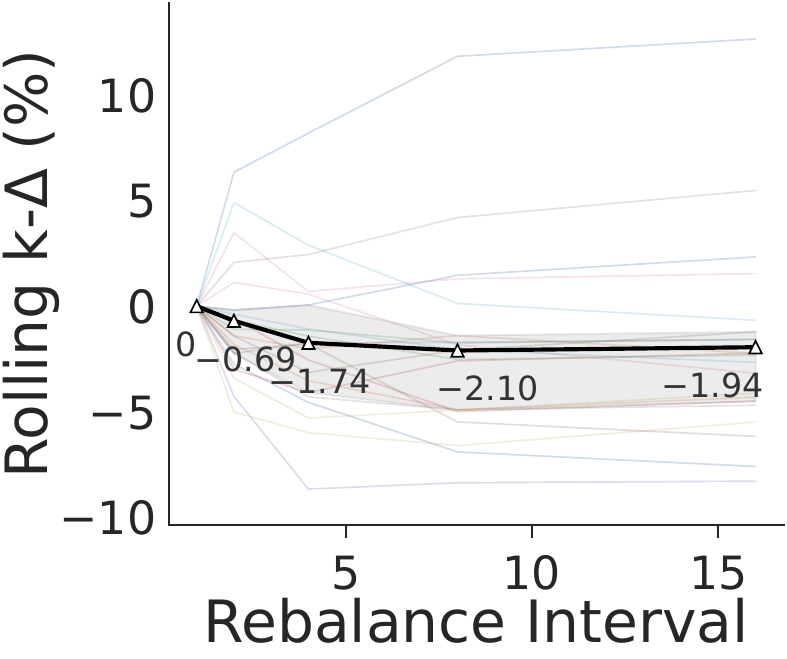}
        \caption{\textbf{Rolling $k$-delta analysis on Polymarket.} We also evaluate rebalance intervals $k \in \{1, 2, 4, 8, 16\}$. The black line denotes the mean performance across 21 models, and the shaded gray region indicates the 25–75\% confidence interval.}
        \label{fig:delta_Polymarket}
    \end{minipage}
    \hfill
    \begin{minipage}[t]{0.32\linewidth}
        \centering
        \includegraphics[width=\linewidth]{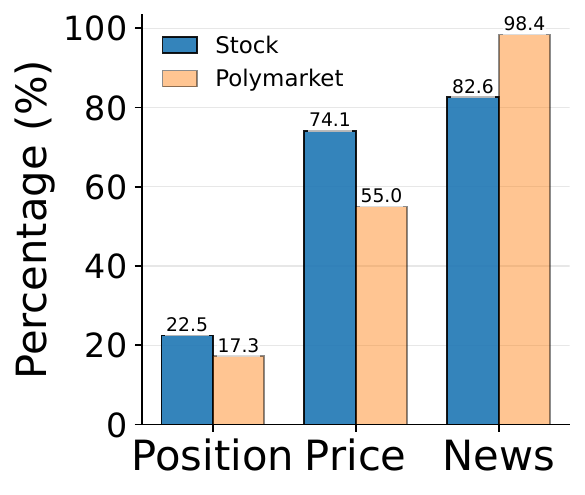}
        \caption{\textbf{Decision-making rationale analysis}. Each bar indicates the proportion of reasoning traces that reference position, price, or news information. A single reasoning trace may include multiple information sources.}
        \label{fig:reason_analysis}
    \end{minipage}
\end{figure}

In this section, we quantitatively analyze two core questions that probe the fundamental capabilities of LLM-based trading agents.
(1) \textit{Are LLM-based agents merely random guessers?} — This examines whether the agents’ trading behaviors reflect meaningful market understanding or simply random fluctuations.
(2) \textit{How do agents reason and make trading decisions?} — This investigates the internal rationale behind their actions, revealing whether their decisions are grounded in coherent reasoning patterns. Together, these analyses shed light on both the effectiveness and interpretability of LLM-based agents in dynamic, uncertain market environments.

\paragraph{Are LLM agents just random guessing?}
To verify that LLM-based trading agents exhibit genuine market awareness rather than random behavior, we design the \textbf{rolling-$k$ delta} ($\Delta_k$) analysis. The key idea is that if the agents’ decisions are random, delaying their actions by several days should not systematically affect performance. Conversely, if they truly adapt to changing market conditions, stale decisions should lead to measurable degradation.
For each trading day $t$, we fix the portfolio position to the one taken $k$ days earlier, $\mathbf{q}^{(k)}_t = \mathbf{q}_{t-k}$, and compute daily and cumulative returns as
\begin{equation}
r^{(k)}_t =
\frac{(\mathbf{q}_{t-k})^\top(\mathbf{p}_{t+1}-\mathbf{p}_t)}{(\mathbf{q}_{t-k})^\top\mathbf{p}_t},
\qquad
CR^{(k)} = \prod_{t=k}^{T-k-1}(1 + r^{(k)}_t) - 1.
\end{equation}
The rolling-$k$ delta is then defined as $\Delta_k = CR^{(k)} - CR^{(0)}$, capturing the cumulative return loss when the agent’s actions lag behind the market by $k$ days.
A negative $\Delta_k$ indicates that more frequent rebalancing improves performance.
As shown in Figure~\ref{fig:delta_stock} and Figure~\ref{fig:delta_Polymarket}, larger $k$ (slower updates) leads to higher degradation, confirming that timely decision updates are beneficial.
Interestingly, in the stock market, returns slightly improve to 0.03\% when $k=2$, suggesting smoother dynamics and lower time sensitivity compared to the Polymarket, where performance degrades 2\% as $k$ increases.
Overall, these results demonstrate that LLM-based agents do not act randomly—their trading strategies depend on contemporaneous market signals, and delaying their actions systematically harms performance.

\paragraph{How do LLM agents reason and make decisions?}
To investigate decision-making rationale, we employ LLM-based reasoning annotation.
For each day’s reasoning trace, another LLM automatically identifies whether the agent’s explanation references:
(1) portfolio position, (2) market price history, or (3) market news.
Figure~\ref{fig:reason_analysis} summarizes the distribution of these factors.
In both markets, \textit{news} emerges as the most frequently cited factor, followed by \textit{market price history}, while \textit{position} information is less dominant.
Moreover, the Polymarket agents rely more heavily on news signals, while stock-market agents emphasize price trends—validating our hypothesis that the two markets exhibit distinct dynamics.
Since the total percentage of reasoning references exceeds 100\%, many decisions integrate multiple information sources, indicating complex reasoning processes.
Specifically, agents often mention “price momentum” when analyzing price history and focus on potential outcomes or implications when discussing market news.

\section{Case Study}
In this section, we show examples of both the U.S stock and Polymarket prediction markets by selecting representative assets among each of them and highlighting two distinct and extreme time points for each to analyze. This allows us to examine the reasoning behind the agents’ decisions and understand how their choices correlate with market conditions.

\subsection{Cash Asset Dynamics in the U.S. Stock Market}

\begin{figure}
    \centering
    \includegraphics[width=\linewidth]{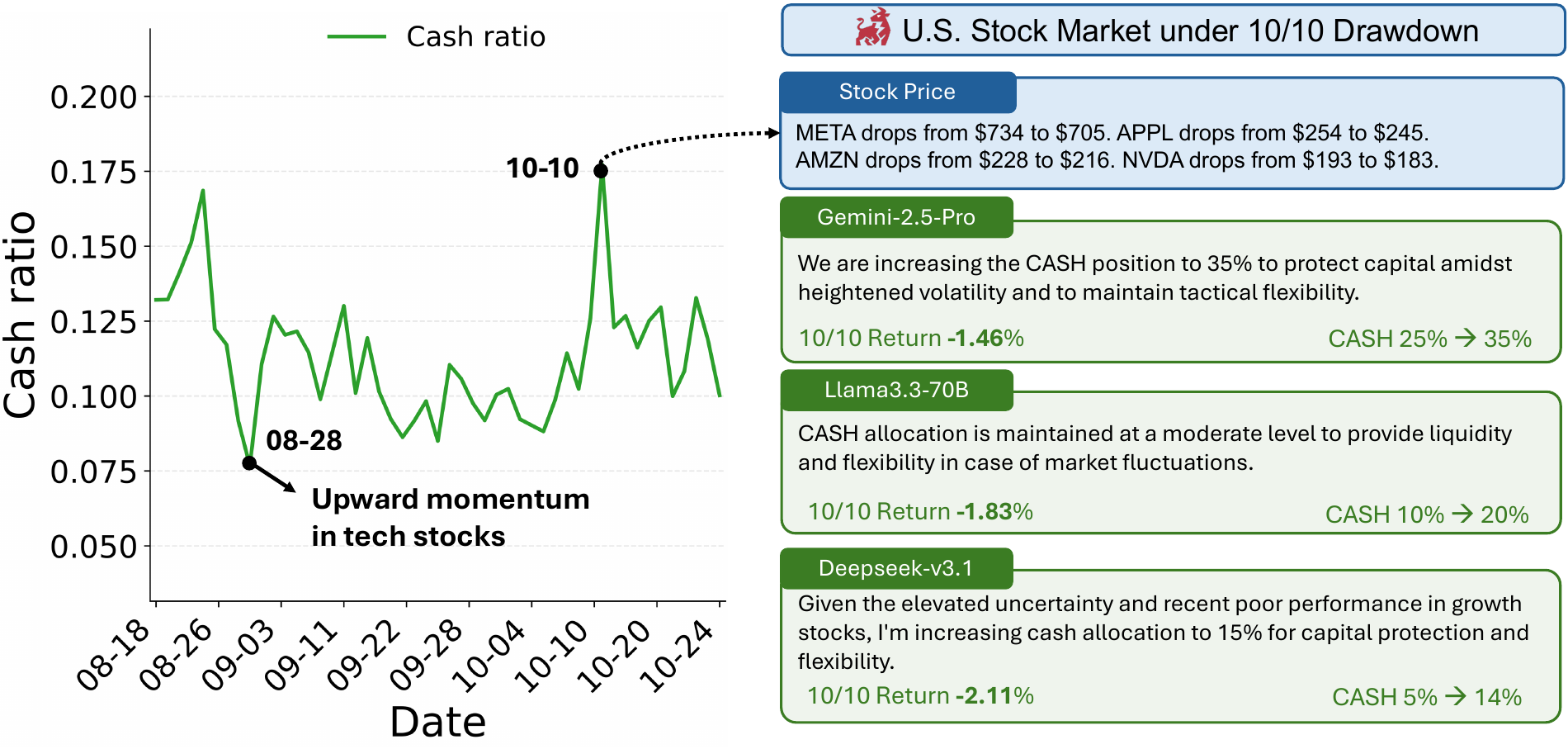}
    \caption{\textbf{Case study for U.S. stock markets.} (\textbf{Left}) The average cash ratio across 21 models over 50 trading days. (\textbf{Right}) A zoomed-in view of the sharp drawdown on October 10, during which portfolios exhibited a sudden increase in cash holdings. We visualize the market condition (price change) and present the reasoning traces of the best-performing model on October 10 (Gemini-2.5-Pro) and one of the worst-performing models on October 10 (DeepSeek-V3.1) for comparison.}
    \label{fig:stock_case_study}
\end{figure}

In Figure~\ref{fig:stock_case_study}, we analyze the dynamics of cash assets in the portfolio and highlight two contrasting market scenarios—a positive (bullish) case and a negative (bearish) case—in the U.S. stock market. The cash ratio serves as an informative indicator of risk management: a higher cash ratio typically reflects greater risk aversion and a defensive stance, whereas a lower cash ratio suggests stronger market confidence and a more aggressive investment posture.

\paragraph{Tech stock rally on August 28}
On August 28, the average cash ratio across all 21 models declined steadily over three consecutive days, dropping from 17\% to 7.5\% within four trading days. This trend coincided with a strong rally in major technology stocks such as Meta (META), Apple (AAPL), and Microsoft (MSFT), which encouraged agents to invest more aggressively and reduce their cash holdings.
Notably, {GPT-4.1}, the agent achieving the highest cumulative return rate, provided the following rationale for its allocation decision:
\textit{“This allocation increases exposure to leading AI and tech growth stocks (NVDA, MSFT, META) following strong earnings momentum and positive analyst sentiment, while maintaining solid positions in diversified blue chips for stability and sector balance.”}
This reasoning explicitly aligns with the observed decrease in the cash ratio during the bullish market phase.

\paragraph{Market drawdown on October 10}
In contrast, the sharp market drawdown on October 10 induced the opposite behavior. Most of the stock prices dropped significantly—Tesla (TSLA) fell over 5\%, while Amazon (AMZN) and Nvidia (NVDA) declined by more than 4\%—leading to negative returns for all agents on that day. In response, most agents increased their cash holdings to mitigate risk, reflecting a collective shift toward a defensive strategy.
As shown on the right of Figure~\ref{fig:stock_case_study}, multiple agents provided similar reasoning related to “increasing CASH positions to protect against volatility.” Among them, {Gemini-2.5-Pro}, which maintained a relatively high cash position and converted additional assets to cash before the downturn, experienced the smallest loss that day.

\subsection{Russia–Ukraine Ceasefire Market in Polymarket Prediction Market}

\begin{figure}
    \centering
    \includegraphics[width=\linewidth]{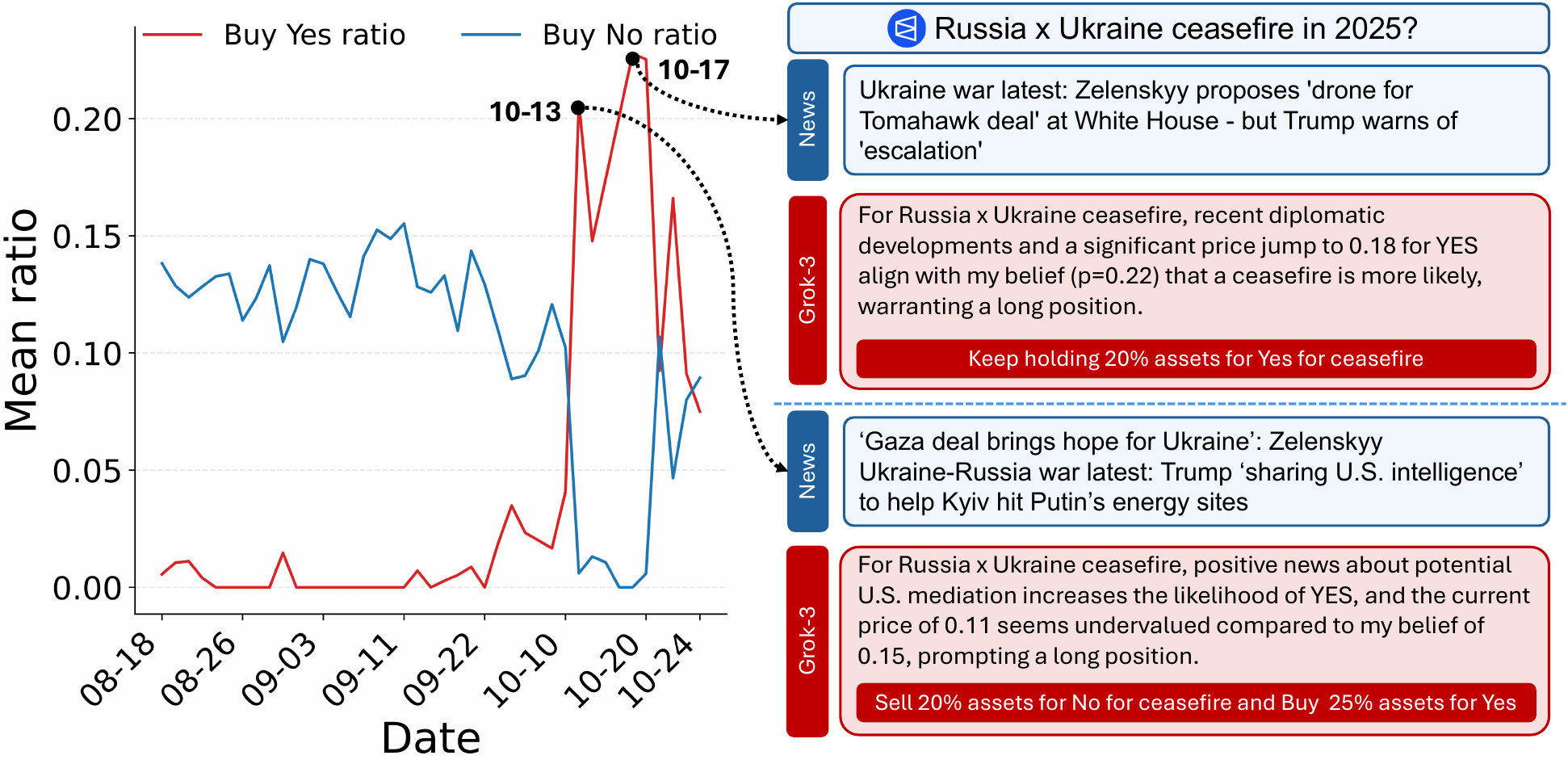}
    \caption{\textbf{Case study in Polymarket prediction markets.} (\textbf{Left}) The average holding ratios of “Yes” and “No” position ratios in the market “Russia × Ukraine ceasefire in 2025?” across 21 models. (\textbf{Right}) A zoomed-in view of two abrupt shifts (October 13 and October 17), along with the corresponding news events and the reasoning traces of Grok-3 explaining these allocation decisions.}
    \label{fig:polymarket_case_study}
\end{figure}

We analyze the market \textit{``Russia × Ukraine ceasefire in 2025?''} on Polymarket, focusing on how real-time news influences the decision-making of LLM-based agents. Polymarket’s high sensitivity to external information makes it a natural testbed to evaluate how models interpret and act on dynamic geopolitical signals.
We select this market for the case study because it experiences frequent fluctuations during the 50 trading days, resulting in distinct behaviors and returns across models. As shown in Figure~\ref{fig:polymarket_case_study}, the \textit{Grok-3} model is able to conduct belief-based reasoning, adjusting its internal estimate of the ceasefire probability from 0.15 on October 13 to 0.22 on October 17.

\paragraph{Reactive change on October 13 without profit}
On October 13, most agents abruptly switched their portfolios from \textit{No} to \textit{Yes} positions after two optimistic news events: (1) \textit{Zelenskyy stated that “the Gaza deal brings hope for Ukraine,”} and (2) reports surfaced that \textit{Trump shared U.S. intelligence to help Kyiv strike Russian energy targets.} These headlines appeared relevant to the ceasefire, prompting agents to buy into the \textit{Yes} position.
However, the Polymarket price showed little actual movement, and this reactive change produced no profit. The news turned out to have limited causal impact on the ceasefire likelihood. This case highlights a key challenge for LLM-based agents: distinguishing between attention-grabbing but non-decisive news and genuinely influential events. Acting on superficial correlations can lead to overreaction and unprofitable trades.

\paragraph{Strategic hold on October 17 with profit}
In contrast, on October 17, when news broke that \textit{Zelenskyy visited the White House}, most agents strengthened their \textit{Yes} positions and held them through the following day. This time, the Polymarket price steadily increased from October 17 to 18, leading to tangible profits.
Unlike the earlier overreaction, agents displayed more grounded reasoning—citing “recent diplomatic developments” and recognizing “a significant price jump to 0.18” as confirming evidence. This scenario illustrates that maintaining positions through credible, high-impact events can yield better outcomes than frequent reactive shifts based on weak signals.

\vspace{-2mm}
\section{Conclusion}

In this work, we present \modelname, a live multi-market environment for evaluating LLM-based agents in realistic portfolio management tasks. \modelname\ introduces a new paradigm for assessing model intelligence beyond static benchmarks, enabling continuous interaction, reasoning, and adaptation within real-time stock and prediction markets. Through 50-day live experiments, we find that strong performance in one market does not generalize to others, underscoring the heterogeneity and specialization required across market types. Moreover, high scores on general-purpose benchmarks like LMArena do not necessarily translate into superior trading performance, highlighting a gap between text intelligence and dynamic decision-making. Finally, our analyses reveal that LLM-based agents rely jointly on historical price trends, market news, allocation history, exhibiting distinct behavioral patterns under extreme conditions. Overall, \modelname\ provides a foundation for studying how LLM-based agents perceive, reason, and act under uncertainty in live and realistic trading environments—paving the way for developing more adaptive, financially grounded, and socially intelligent agent systems.

\section*{Limitation and Future Work}

Despite demonstrating the feasibility of evaluating LLM-based trading agents in live multi-market environments, our framework still has thee main limitations that point to promising directions for future research and development.

\paragraph{Transaction costs and market frictions}
Our current environment and evaluation do not account for transaction fees, bid–ask spreads, liquidity constraints, or other real-world trading frictions. Ignoring these factors may overestimate achievable returns, especially for strategies that rely on frequent rebalancing. Future work will incorporate more realistic cost models and slippage simulations to better approximate real trading conditions.

\paragraph{Limited observation and action space}
The current framework constrains both the observation and action spaces due to the limited context length of existing LLMs.
For the \emph{observation space}, the agent can only access a restricted temporal window of price, position, and news histories, and these are limited to a small set of markets in both the stock market and the prediction market.
Moreover, news inputs are truncated to titles and abstracts rather than full articles, preventing the agent from incorporating long-form textual information that may contain deeper market signals.
For the \emph{action space}, the scope of possible trading actions is similarly constrained by the limited number of supported markets, reducing the complexity and richness of allocation decisions.
Future work could extend the framework to support longer temporal horizons, richer textual context, and dynamic market expansion—enabling agents to observe and act within more realistic, information-rich environments.

\paragraph{Simplified agent design}
The current agent architecture integrates basic tool use and memory under the ReAct framework but remains limited in reasoning depth and temporal abstraction. Future work can enhance each component systematically. For \textit{tool use}, agents can be equipped with more specialized analytical and retrieval tools for financial reasoning, news interpretation, and risk assessment. For \textit{memory}, richer hierarchical and long-term memory mechanisms can be introduced to capture temporal dependencies and retain cross-market knowledge over extended horizons. Beyond the basic ReAct-based setups, the current framework can be extended to a \textit{multi-agent} paradigm, such as TradingAgents~\citep{xiao2024tradingagents}, to better model heterogeneous roles and market interactions. Finally, incorporating reinforcement learning (RL) to train trading agents~\citep{xiao2025trading} represents a promising direction for improving decision quality—enabling agents to learn from experience, refine their reasoning, and continuously adapt to dynamic market conditions.

\section*{Open-source Application}
To democratize research on LLM-based trading agents, we release an open-source Python package, \texttt{live-trade-bench}\footnote{\url{https://pypi.org/project/live-trade-bench/}}
, which provides simple APIs for data collection, environment setup, and agent construction.
Building on this package, we also develop a web application that deploys our trading environment in real time, enabling live data streaming and interactive monitoring of agent performance.
Details of the user interface (UI) design are provided in Appendix~\S\ref{ui-details}.

\clearpage
\newpage

\bibliographystyle{abbrvnat}
\bibliography{references}

@article{Zhang2024AMFA,
  title={A Multimodal Foundation Agent for Financial Trading: Tool-Augmented, Diversified, and Generalist},
  author={Wentao Zhang and Lingxuan Zhao and Haochong Xia and Shuo Sun and Jiaze Sun and Molei Qin and Xinyi Li and Yuqing Zhao and Yilei Zhao and Xinyu Cai and Longtao Zheng and Xinrun Wang and Bo An},
  journal={Proceedings of the 30th ACM SIGKDD Conference on Knowledge Discovery and Data Mining},
  year={2024},
  url={http://dl.acm.org/citation.cfm?id=3671801}
}

@inproceedings{chiang2024chatbot,
  title={Chatbot arena: An open platform for evaluating llms by human preference},
  author={Chiang, Wei-Lin and Zheng, Lianmin and Sheng, Ying and Angelopoulos, Anastasios Nikolas and Li, Tianle and Li, Dacheng and Zhu, Banghua and Zhang, Hao and Jordan, Michael and Gonzalez, Joseph E and others},
  booktitle={Forty-first International Conference on Machine Learning},
  year={2024}
}

@article{Yu2024FinConASA,
  title={FinCon: A Synthesized LLM Multi-Agent System with Conceptual Verbal Reinforcement for Enhanced Financial Decision Making},
  author={Yangyang Yu and Zhiyuan Yao and Haohang Li and Zhiyang Deng and Yupeng Cao and Zhi Chen and Jordan W. Suchow and Rong Liu and Zhenyu Cui and Denghui Zhang and K. Subbalakshmi and Guojun Xiong and Yueru He and Jimin Huang and Dong Li and Qianqian Xie},
  journal={ArXiv},
  year={2024},
  volume={abs/2407.06567},
  url={https://api.semanticscholar.org/CorpusId:271064881}
}

@article{Gu2024AdaptiveAEA,
  title={Adaptive and Explainable Margin Trading via Large Language Models on Portfolio Management},
  author={Jingyi Gu and J. Ye and Guiling Wang and Wenpeng Yin},
  journal={Proceedings of the 5th ACM International Conference on AI in Finance},
  year={2024},
  url={https://api.semanticscholar.org/CorpusId:274086023}
}

@article{Kou2024AutomateSFA,
  title={Automate Strategy Finding with LLM in Quant investment},
  author={Zhizhuo Kou and Holam Yu and Jingshu Peng and Lei Chen},
  journal={ArXiv},
  year={2024},
  volume={abs/2409.06289},
  url={https://api.semanticscholar.org/CorpusId:272550540}
}

@article{Han2023SelectATA,
  title={Select and Trade: Towards Unified Pair Trading with Hierarchical Reinforcement Learning},
  author={Weiguang Han and Boyi Zhang and Qianqian Xie and Min Peng and Yanzhao Lai and Jimin Huang},
  journal={Proceedings of the 29th ACM SIGKDD Conference on Knowledge Discovery and Data Mining},
  year={2023},
  url={https://api.semanticscholar.org/CorpusId:256231505}
}

@article{Han2023MasteringPTA,
  title={Mastering Pair Trading with Risk-Aware Recurrent Reinforcement Learning},
  author={Weiguang Han and Jimin Huang and Qianqian Xie and Boyi Zhang and Yanzhao Lai and Min Peng},
  journal={ArXiv},
  year={2023},
  volume={abs/2304.00364},
  url={https://api.semanticscholar.org/CorpusId:257913550}
}

@article{Briola2021DeepRLA,
  title={Deep Reinforcement Learning for Active High Frequency Trading},
  author={Antonio Briola and J. Turiel and Riccardo Marcaccioli and T. Aste},
  journal={ArXiv},
  year={2021},
  volume={abs/2101.07107},
  url={https://api.semanticscholar.org/CorpusId:231632086}
}

@inproceedings{Ma2025AgentTAA,
  title={Agent Trading Arena: A Study on Numerical Understanding in LLM-Based Agents},
  author={Tianmi Ma and Jiawei Du and Wenxin Huang and Wenjie Wang and Liang Xie and Xian Zhong and Joey Tianyi Zhou},
  booktitle={unknown},
  year={2025},
  url={https://api.semanticscholar.org/CorpusId:276580341}
}

@inproceedings{Li2025WillLBA,
  title={Will LLMs be Professional at Fund Investment? DeepFund: A Live Arena Perspective},
  author={Changlun Li and Yao Shi and Yuyu Luo and Nan Tang},
  booktitle={unknown},
  year={2025},
  url={https://api.semanticscholar.org/CorpusId:277272186}
}

@article{Li2025TimeTIA,
  title={Time Travel is Cheating: Going Live with DeepFund for Real-Time Fund Investment Benchmarking},
  author={Changlun Li and Yao Shi and Chen Wang and Qiqi Duan and Runke Ruan and Weijie Huang and Haonan Long and Lijun Huang and Yuyu Luo and Nan Tang},
  journal={ArXiv},
  year={2025},
  volume={abs/2505.11065},
  url={https://api.semanticscholar.org/CorpusId:278715123}
}

@article{Li2024INVESTORBENCHABA,
  title={INVESTORBENCH: A Benchmark for Financial Decision-Making Tasks with LLM-based Agent},
  author={Haohang Li and Yupeng Cao and Yangyang Yu and Shashidhar Reddy Javaji and Zhiyang Deng and Yueru He and Yuechen Jiang and Zining Zhu and K. Subbalakshmi and Guojun Xiong and Jimin Huang and Lingfei Qian and Xueqing Peng and Qianqian Xie and Jordan W. Suchow},
  journal={ArXiv},
  year={2024},
  volume={abs/2412.18174},
  url={https://api.semanticscholar.org/CorpusId:274992211}
}

@article{GarridoMerchan2024DeepRLA,
  title={Deep Reinforcement Learning Agents for Strategic Production Policies in Microeconomic Market Simulations},
  author={Eduardo C. Garrido-Merch'an and Maria Coronado Vaca and {\'A}lvaro L{\'o}pez-L{\'o}pez and Carlos Mart{\'i}nez de Ibarreta},
  journal={ArXiv},
  year={2024},
  volume={abs/2410.20550},
  url={https://api.semanticscholar.org/CorpusId:273653990}
}

@article{zhou2023sotopia,
  title={Sotopia: Interactive evaluation for social intelligence in language agents},
  author={Zhou, Xuhui and Zhu, Hao and Mathur, Leena and Zhang, Ruohong and Yu, Haofei and Qi, Zhengyang and Morency, Louis-Philippe and Bisk, Yonatan and Fried, Daniel and Neubig, Graham and others},
  journal={arXiv preprint arXiv:2310.11667},
  year={2023}
}

@article{jimenez2023swe,
  title={Swe-bench: Can language models resolve real-world github issues?},
  author={Jimenez, Carlos E and Yang, John and Wettig, Alexander and Yao, Shunyu and Pei, Kexin and Press, Ofir and Narasimhan, Karthik},
  journal={arXiv preprint arXiv:2310.06770},
  year={2023}
}

@article{zhou2023webarena,
  title={Webarena: A realistic web environment for building autonomous agents},
  author={Zhou, Shuyan and Xu, Frank F and Zhu, Hao and Zhou, Xuhui and Lo, Robert and Sridhar, Abishek and Cheng, Xianyi and Ou, Tianyue and Bisk, Yonatan and Fried, Daniel and others},
  journal={arXiv preprint arXiv:2307.13854},
  year={2023}
}

@article{Aksitov2023ReSTMRA,
  title={ReST meets ReAct: Self-Improvement for Multi-Step Reasoning LLM Agent},
  author={Renat Aksitov and Sobhan Miryoosefi and Zong-xiao Li and Daliang Li and Sheila Babayan and Kavya Kopparapu and Zachary Fisher and Ruiqi Guo and Sushant Prakash and Pranesh Srinivasan and M. Zaheer and Felix X. Yu and Sanjiv Kumar},
  journal={ArXiv},
  year={2023},
  volume={abs/2312.10003},
  url={https://api.semanticscholar.org/CorpusId:266335848}
}

@article{Putta2024AgentQAA,
  title={Agent Q: Advanced Reasoning and Learning for Autonomous AI Agents},
  author={Pranav Putta and Edmund Mills and Naman Garg and S. Motwani and Chelsea Finn and Divyansh Garg and Rafael Rafailov},
  journal={ArXiv},
  year={2024},
  volume={abs/2408.07199},
  url={https://api.semanticscholar.org/CorpusId:271865516}
}

@article{bigeard2025finance,
  title={Finance Agent Benchmark: Benchmarking LLMs on Real-world Financial Research Tasks},
  author={Bigeard, Antoine and Nashold, Langston and Krishnan, Rayan and Wu, Shirley},
  journal={arXiv preprint arXiv:2508.00828},
  year={2025}
}

@article{Koh2024TreeSFA,
  title={Tree Search for Language Model Agents},
  author={Jing Yu Koh and Stephen McAleer and Daniel Fried and Ruslan Salakhutdinov},
  journal={ArXiv},
  year={2024},
  volume={abs/2407.01476},
  url={https://api.semanticscholar.org/CorpusId:270870063}
}

@article{he2024webvoyager,
  title={Webvoyager: Building an end-to-end web agent with large multimodal models},
  author={He, Hongliang and Yao, Wenlin and Ma, Kaixin and Yu, Wenhao and Dai, Yong and Zhang, Hongming and Lan, Zhenzhong and Yu, Dong},
  journal={arXiv preprint arXiv:2401.13919},
  year={2024}
}

@article{xie2024osworld,
  title={Osworld: Benchmarking multimodal agents for open-ended tasks in real computer environments},
  author={Xie, Tianbao and Zhang, Danyang and Chen, Jixuan and Li, Xiaochuan and Zhao, Siheng and Cao, Ruisheng and Hua, Toh J and Cheng, Zhoujun and Shin, Dongchan and Lei, Fangyu and others},
  journal={Advances in Neural Information Processing Systems},
  volume={37},
  pages={52040--52094},
  year={2024}
}

@article{koh2024visualwebarena,
  title={Visualwebarena: Evaluating multimodal agents on realistic visual web tasks},
  author={Koh, Jing Yu and Lo, Robert and Jang, Lawrence and Duvvur, Vikram and Lim, Ming Chong and Huang, Po-Yu and Neubig, Graham and Zhou, Shuyan and Salakhutdinov, Ruslan and Fried, Daniel},
  journal={arXiv preprint arXiv:2401.13649},
  year={2024}
}

@article{jiang2023followbench,
  title={Followbench: A multi-level fine-grained constraints following benchmark for large language models},
  author={Jiang, Yuxin and Wang, Yufei and Zeng, Xingshan and Zhong, Wanjun and Li, Liangyou and Mi, Fei and Shang, Lifeng and Jiang, Xin and Liu, Qun and Wang, Wei},
  journal={arXiv preprint arXiv:2310.20410},
  year={2023}
}

@article{cobbe2021training,
  title={Training verifiers to solve math word problems},
  author={Cobbe, Karl and Kosaraju, Vineet and Bavarian, Mohammad and Chen, Mark and Jun, Heewoo and Kaiser, Lukasz and Plappert, Matthias and Tworek, Jerry and Hilton, Jacob and Nakano, Reiichiro and others},
  journal={arXiv preprint arXiv:2110.14168},
  year={2021}
}

@article{zhou2023instruction,
  title={Instruction-following evaluation for large language models},
  author={Zhou, Jeffrey and Lu, Tianjian and Mishra, Swaroop and Brahma, Siddhartha and Basu, Sujoy and Luan, Yi and Zhou, Denny and Hou, Le},
  journal={arXiv preprint arXiv:2311.07911},
  year={2023}
}

@misc{pyatkin2025generalizing,
   title={Generalizing Verifiable Instruction Following}, 
   author={Valentina Pyatkin and Saumya Malik and Victoria Graf and Hamish Ivison and Shengyi Huang and Pradeep Dasigi and Nathan Lambert and Hannaneh Hajishirzi},
   year={2025},
   eprint={TODO},
   archivePrefix={arXiv},
   primaryClass={cs.CL}
}

@article{quan2025codeelo,
  title={Codeelo: Benchmarking competition-level code generation of llms with human-comparable elo ratings},
  author={Quan, Shanghaoran and Yang, Jiaxi and Yu, Bowen and Zheng, Bo and Liu, Dayiheng and Yang, An and Ren, Xuancheng and Gao, Bofei and Miao, Yibo and Feng, Yunlong and others},
  journal={arXiv preprint arXiv:2501.01257},
  year={2025}
}

@misc{2023opencompass,
    title={OpenCompass: A Universal Evaluation Platform for Foundation Models},
    author={OpenCompass Contributors},
    howpublished = {\url{https://github.com/open-compass/opencompass}},
    year={2023}
}

@inproceedings{rein2024gpqa,
  title={Gpqa: A graduate-level google-proof q\&a benchmark},
  author={Rein, David and Hou, Betty Li and Stickland, Asa Cooper and Petty, Jackson and Pang, Richard Yuanzhe and Dirani, Julien and Michael, Julian and Bowman, Samuel R},
  booktitle={First Conference on Language Modeling},
  year={2024}
}

@article{phan2025humanity,
  title={Humanity's last exam},
  author={Phan, Long and Gatti, Alice and Han, Ziwen and Li, Nathaniel and Hu, Josephina and Zhang, Hugh and Zhang, Chen Bo Calvin and Shaaban, Mohamed and Ling, John and Shi, Sean and others},
  journal={arXiv preprint arXiv:2501.14249},
  year={2025}
}

@article{hendrycks2020measuring,
  title={Measuring massive multitask language understanding},
  author={Hendrycks, Dan and Burns, Collin and Basart, Steven and Zou, Andy and Mazeika, Mantas and Song, Dawn and Steinhardt, Jacob},
  journal={arXiv preprint arXiv:2009.03300},
  year={2020}
}

@misc{nie2025uqassessinglanguagemodels,
  title={UQ: Assessing Language Models on Unsolved Questions}, 
  author={Fan Nie and Ken Ziyu Liu and Zihao Wang and Rui Sun and Wei Liu and Weijia Shi and Huaxiu Yao and Linjun Zhang and Andrew Y. Ng and James Zou and Sanmi Koyejo and Yejin Choi and Percy Liang and Niklas Muennighoff},
  year={2025},
  eprint={2508.17580},
  archivePrefix={arXiv},
  primaryClass={cs.CL},
  url={https://arxiv.org/abs/2508.17580}
}

@article{Kasai2022RealTimeQWA,
  title={RealTime QA: What's the Answer Right Now?},
  author={Jungo Kasai and Keisuke Sakaguchi and Yoichi Takahashi and Ronan Le Bras and Akari Asai and Xinyan Velocity Yu and Dragomir R. Radev and Noah A. Smith and Yejin Choi and Kentaro Inui},
  journal={ArXiv},
  year={2022},
  volume={abs/2207.13332},
  url={https://api.semanticscholar.org/CorpusId:251105205}
}

@article{team2025kimi,
  title={Kimi k2: Open agentic intelligence},
  author={Team, Kimi and Bai, Yifan and Bao, Yiping and Chen, Guanduo and Chen, Jiahao and Chen, Ningxin and Chen, Ruijue and Chen, Yanru and Chen, Yuankun and Chen, Yutian and others},
  journal={arXiv preprint arXiv:2507.20534},
  year={2025}
}

@article{guo2025deepseek,
  title={Deepseek-r1: Incentivizing reasoning capability in llms via reinforcement learning},
  author={Guo, Daya and Yang, Dejian and Zhang, Haowei and Song, Junxiao and Zhang, Ruoyu and Xu, Runxin and Zhu, Qihao and Ma, Shirong and Wang, Peiyi and Bi, Xiao and others},
  journal={arXiv preprint arXiv:2501.12948},
  year={2025}
}

@article{yang2025qwen3,
  title={Qwen3 technical report},
  author={Yang, An and Li, Anfeng and Yang, Baosong and Zhang, Beichen and Hui, Binyuan and Zheng, Bo and Yu, Bowen and Gao, Chang and Huang, Chengen and Lv, Chenxu and others},
  journal={arXiv preprint arXiv:2505.09388},
  year={2025}
}

@article{xiao2024tradingagents,
  title={TradingAgents: Multi-agents LLM financial trading framework},
  author={Xiao, Yijia and Sun, Edward and Luo, Di and Wang, Wei},
  journal={arXiv preprint arXiv:2412.20138},
  year={2024}
}

@article{xiao2025trading,
  title={Trading-r1: Financial trading with llm reasoning via reinforcement learning},
  author={Xiao, Yijia and Sun, Edward and Chen, Tong and Wu, Fang and Luo, Di and Wang, Wei},
  journal={arXiv preprint arXiv:2509.11420},
  year={2025}
}

@article{wei2022chain,
  title={Chain-of-thought prompting elicits reasoning in large language models},
  author={Wei, Jason and Wang, Xuezhi and Schuurmans, Dale and Bosma, Maarten and Xia, Fei and Chi, Ed and Le, Quoc V and Zhou, Denny and others},
  journal={Advances in neural information processing systems},
  volume={35},
  pages={24824--24837},
  year={2022}
}

@inproceedings{LopezLira2025CanLLA,
  title={Can Large Language Models Trade? Testing Financial Theories with LLM Agents in Market Simulations},
  author={Alejandro Lopez-Lira},
  booktitle={unknown},
  year={2025},
  url={https://api.semanticscholar.org/CorpusId:277787336}
}

@article{Chen2023PutYMA,
  title={Put Your Money Where Your Mouth Is: Evaluating Strategic Planning and Execution of LLM Agents in an Auction Arena},
  author={Jiangjie Chen and Siyu Yuan and Rong Ye and Bodhisattwa Prasad Majumder and Kyle Richardson},
  journal={ArXiv},
  year={2023},
  volume={abs/2310.05746},
  url={https://api.semanticscholar.org/CorpusId:263831697}
}

@article{Zhang2024WhenAMA,
  title={When AI Meets Finance (StockAgent): Large Language Model-based Stock Trading in Simulated Real-world Environments},
  author={Chong Zhang and Xinyi Liu and Mingyu Jin and Zhongmou Zhang and Lingyao Li and Zhengting Wang and Wenyue Hua and Dong Shu and Suiyuan Zhu and Xiaobo Jin and Sujian Li and Mengnan Du and Yongfeng Zhang},
  journal={ArXiv},
  year={2024},
  volume={abs/2407.18957},
  url={https://api.semanticscholar.org/CorpusId:271533952}
}

@article{Emmanoulopoulos2025ToTOA,
  title={To Trade or Not to Trade: An Agentic Approach to Estimating Market Risk Improves Trading Decisions},
  author={Dimitrios Emmanoulopoulos and Ollie Olby and Justin Lyon and Namid R Stillman},
  journal={ArXiv},
  year={2025},
  volume={abs/2507.08584},
  url={https://api.semanticscholar.org/CorpusId:280281425}
}

@inproceedings{Li2025ProfitMRA,
  title={Profit Mirage: Revisiting Information Leakage in LLM-based Financial Agents},
  author={Xiangyu Li and Yawen Zeng and Xiaofen Xing and Jin Xu and Xiangmin Xu},
  booktitle={unknown},
  year={2025},
  url={https://arxiv.org/pdf/2510.07920.pdf}
}

@article{xu2024tradingagent,
  title={TradingAgent: Multi-Agent Financial Decision-Making with LLMs},
  author={Xu, Tianyang and Li, Haoran and Liu, Jie},
  journal={arXiv preprint arXiv:2412.20138},
  year={2024}
}

@article{Tang2025AlphaAgentLAA,
  title={AlphaAgent: LLM-Driven Alpha Mining with Regularized Exploration to Counteract Alpha Decay},
  author={Ziyi Tang and Zechuan Chen and Jiarui Yang and Jiayao Mai and Yongsen Zheng and Keze Wang and Jinrui Chen and Liang Lin},
  journal={ArXiv},
  year={2025},
  volume={abs/2502.16789},
  url={https://api.semanticscholar.org/CorpusId:276574595}
}

@article{Tian2025TradingGroupAMA,
  title={TradingGroup: A Multi-Agent Trading System with Self-Reflection and Data-Synthesis},
  author={Feng Tian and Flora D. Salim and Hao Xue},
  journal={ArXiv},
  year={2025},
  volume={abs/2508.17565},
  url={https://api.semanticscholar.org/CorpusId:280711571}
}

@article{Xiao2024TradingAgentsMLA,
  title={TradingAgents: Multi-Agents LLM Financial Trading Framework},
  author={Yijia Xiao and Edward Sun and Di Luo and Wei Wang},
  journal={ArXiv},
  year={2024},
  volume={abs/2412.20138},
  url={https://api.semanticscholar.org/CorpusId:275133732}
}

@article{Li2025CanLFA,
  title={Can LLM-based Financial Investing Strategies Outperform the Market in Long Run?},
  author={Weixian Waylon Li and Hyeonjun Kim and Mihai Cucuringu and Tiejun Ma},
  journal={ArXiv},
  year={2025},
  volume={abs/2505.07078},
  url={https://api.semanticscholar.org/CorpusId:278501425}
}

@article{Lucarelli2020ADQA,
  title={A deep Q-learning portfolio management framework for the cryptocurrency market},
  author={Giorgio Lucarelli and M. Borrotti},
  journal={Neural Computing and Applications},
  year={2020},
  volume={32},
  pages={17229 - 17244},
  url={https://doi.org/10.1007/s00521-020-05359-8}
}

@article{Sun2021ReinforcementLFA,
  title={Reinforcement Learning for Quantitative Trading},
  author={Shuo Sun and R. Wang and Bo An},
  journal={ACM Transactions on Intelligent Systems and Technology},
  year={2021},
  volume={14},
  pages={1 - 29},
  url={https://api.semanticscholar.org/CorpusId:238198196}
}

@article{Ye2020ReinforcementLearningBPA,
  title={Reinforcement-Learning based Portfolio Management with Augmented Asset Movement Prediction States},
  author={Yunan Ye and Hengzhi Pei and Boxin Wang and Pin-Yu Chen and Yada Zhu and Jun Xiao and Bo Li},
  journal={ArXiv},
  year={2020},
  volume={abs/2002.05780},
  url={https://arxiv.org/pdf/2002.05780.pdf}
}

@article{Jumadinova2011AMPA,
  title={A multi-agent prediction market based on partially observable stochastic game},
  author={Janyl Jumadinova and P. Dasgupta},
  journal={ArXiv},
  year={2011},
  volume={abs/1203.6035},
  url={https://api.semanticscholar.org/CorpusId:14232149}
}

@inproceedings{DeHaven2024MinutebyMinuteFMA,
  title={Minute-by-Minute: Financial Markets' Reaction to the 2020 U.S. Election},
  author={Matthew DeHaven and Hannah Firestone and Chris Webster},
  booktitle={unknown},
  year={2024},
  url={https://api.semanticscholar.org/CorpusId:271038710}
}

@article{Koning2022BettingMEA,
  title={Betting market efficiency and prediction in binary choice models},
  author={R. Koning and Renske Zijm},
  journal={Annals of Operations Research},
  year={2022},
  volume={325},
  pages={135 - 148},
  url={https://api.semanticscholar.org/CorpusId:248463684}
}

@article{Heinrich2021FactorIAA,
  title={Factor investing: alpha concentration versus diversification},
  author={Lars Heinrich and A. Shivarova and Martin Zurek},
  journal={Journal of Asset Management},
  year={2021},
  volume={22},
  pages={464 - 487},
  url={https://api.semanticscholar.org/CorpusId:236248859}
}

@inproceedings{Sun2024CombiningTBA,
  title={Combining Transformer based Deep Reinforcement Learning with Black-Litterman Model for Portfolio Optimization},
  author={Ruoyu Sun and Angelos Stefanidis and Zhengyong Jiang and Jionglong Su Xi'an Jiaotong-Liverpool University and S. O. Mathematics and Physics and Department of Financial and Actuarial Mathematics Xi'an Jiaotong-Liverpool University En College and AI Schoolof and Ad-hoc Computing},
  booktitle={unknown},
  year={2024},
  url={https://api.semanticscholar.org/CorpusId:268032065}
}

@article{Zhang2020AutoAlphaAEA,
  title={AutoAlpha: an Efficient Hierarchical Evolutionary Algorithm for Mining Alpha Factors in Quantitative Investment},
  author={T. Zhang and Yuanqi Li and Yifei Jin and Jian Li},
  journal={arXiv: Computational Finance},
  year={2020},
  url={https://api.semanticscholar.org/CorpusId:211171490}
}

@article{Islam2025TheEOA,
  title={The Evolution of Alpha in Finance Harnessing Human Insight and LLM Agents},
  author={Mohammad Rubyet Islam},
  journal={ArXiv},
  year={2025},
  volume={abs/2505.14727},
  url={https://api.semanticscholar.org/CorpusId:278782847}
}

@article{Zhang2025FinWorldAAA,
  title={FinWorld: An All-in-One Open-Source Platform for End-to-End Financial AI Research and Deployment},
  author={Wentao Zhang and Yilei Zhao and Chuqiao Zong and Xinrun Wang and Bo An},
  journal={ArXiv},
  year={2025},
  volume={abs/2508.02292},
  url={https://api.semanticscholar.org/CorpusId:280421572}
}

@article{Gao2024SimulatingFMA,
  title={Simulating Financial Market via Large Language Model based Agents},
  author={Shen Gao and Yuntao Wen and Minghang Zhu and Jianing Wei and Yuhan Cheng and Qunzi Zhang and Shuo Shang},
  journal={ArXiv},
  year={2024},
  volume={abs/2406.19966},
  url={https://api.semanticscholar.org/CorpusId:270845937}
}

@article{wang2025tradingr1,
  title={Trading-R1: Reinforcement-Guided Large Language Models for Financial Markets},
  author={Wang, Zhiyu and Liu, Xin and Zhao, Han},
  journal={arXiv preprint arXiv:2509.11420},
  year={2025}
}

@article{li2025contesttrade,
  title={ContestTrade: A Multi-Agent Trading System Based on Internal Contest Mechanism},
  author={Li, Qiang and Zhao, Xinyi},
  journal={arXiv preprint arXiv:2508.00554},
  year={2025}
}

@article{zhang2025tradinggroup,
  title={TradingGroup: Self-Reflective Multi-Agent System for Financial Markets},
  author={Zhang, Yue and Wang, Rui},
  journal={arXiv preprint arXiv:2508.17565},
  year={2025}
}

@article{gao2024finbench,
  title={FinBench: Benchmarking LLMs for Financial Decision-Making},
  author={Gao, Han and Chen, Jie and Li, Xin},
  journal={NeurIPS},
  year={2024}
}

@article{jiang2025finagent,
  title={FinAgent: Evaluating Financial Reasoning in Large Language Models},
  author={Jiang, Zeyu and Zhou, Shuo},
  journal={arXiv preprint arXiv:2506.03142},
  year={2025}
}

@inproceedings{Li2023TradingGPTMSA,
  title={TradingGPT: Multi-Agent System with Layered Memory and Distinct Characters for Enhanced Financial Trading Performance},
  author={Yang Li and Yangyang Yu and Haohang Li and Z. Chen and K. Khashanah},
  year={2023},
  url={https://api.semanticscholar.org/CorpusId:261582775}
}

@inproceedings{yao2022react,
  title={React: Synergizing reasoning and acting in language models},
  author={Yao, Shunyu and Zhao, Jeffrey and Yu, Dian and Du, Nan and Shafran, Izhak and Narasimhan, Karthik R and Cao, Yuan},
  booktitle={The eleventh international conference on learning representations},
  year={2022}
}

@misc{anthropic2024claude3,
  author       = {Anthropic},
  title        = {Claude 3.7 Sonnet and Claude Code},
  year         = {2025},
  howpublished = {\url{https://www.anthropic.com/news/claude-3-7-sonnet}},
}

@misc{llama4,
  author       = {Meta},
  title        = {The Llama 4 herd: The beginning of a new era of natively multimodal AI innovation},
  year         = {2025},
  howpublished = {\url{https://ai.meta.com/blog/llama-4-multimodal-intelligence/}},
}

@misc{llama33,
  author       = {Meta},
  title        = {Llama 3.3},
  year         = {2025},
  howpublished = {\url{https://www.llama.com/docs/model-cards-and-prompt-formats/llama3_3/}},
}

@misc{gpt5,
  author       = {OpenAI},
  title        = {GPT-5 is here},
  year         = {2025},
  howpublished = {\url{https://openai.com/zh-Hans-CN/gpt-5/}},
}

@misc{gpt41,
  author       = {OpenAI},
  title        = {Introducing GPT-4.1 in the API},
  year         = {2025},
  howpublished = {\url{https://openai.com/index/gpt-4-1/}},
}

@misc{gpto3,
  author       = {OpenAI},
  title        = {Introducing OpenAI o3 and o4-mini},
  year         = {2025},
  howpublished = {\url{https://openai.com/zh-Hans-CN/index/introducing-o3-and-o4-mini/}},
}

@misc{deepseek31,
  author       = {DeepSeek},
  title        = {DeepSeek-V3.1 Release},
  year         = {2025},
  howpublished = {\url{https://api-docs.deepseek.com/news/news250821}},
}

@article{Yang2024Qwen25TR,
  title={Qwen2.5 Technical Report},
  author={Qwen An Yang and Baosong Yang and Beichen Zhang and Binyuan Hui and Bo Zheng and Bowen Yu and Chengyuan Li and Dayiheng Liu and Fei Huang and Guanting Dong and Haoran Wei and Huan Lin and Jian Yang and Jianhong Tu and Jianwei Zhang and Jianxin Yang and Jiaxin Yang and Jingren Zhou and Junyang Lin and Kai Dang and Keming Lu and Keqin Bao and Kexin Yang and Le Yu and Mei Li and Mingfeng Xue and Pei Zhang and Qin Zhu and Rui Men and Runji Lin and Tianhao Li and Tingyu Xia and Xingzhang Ren and Xuancheng Ren and Yang Fan and Yang Su and Yi-Chao Zhang and Yunyang Wan and Yuqi Liu and Zeyu Cui and Zhenru Zhang and Zihan Qiu and Shanghaoran Quan and Zekun Wang},
  journal={ArXiv},
  year={2024},
  volume={abs/2412.15115},
  url={https://api.semanticscholar.org/CorpusID:274859421}
}

@misc{anthropic2024claude4,
  author       = {Anthropic},
  title        = {Introducing Claude 4},
  year         = {2025},
  howpublished = {\url{https://www.anthropic.com/news/claude-4}},
}

@misc{anthropic2024claude41,
  author       = {Anthropic},
  title        = {Claude Opus 4.1},
  year         = {2025},
  howpublished = {\url{https://www.anthropic.com/news/claude-opus-4-1}},
}

@misc{grok3,
  author       = {xAI},
  title        = {Grok 3 Beta — The Age of Reasoning Agents},
  year         = {2025},
  howpublished = {\url{https://x.ai/news/grok-3}},
}

@misc{grok4,
  author       = {xAI},
  title        = {Grok 4},
  year         = {2025},
  howpublished = {\url{https://x.ai/news/grok-4}},
}

@article{hurst2024gpt,
  title={Gpt-4o system card},
  author={Hurst, Aaron and Lerer, Adam and Goucher, Adam P and Perelman, Adam and Ramesh, Aditya and Clark, Aidan and Ostrow, AJ and Welihinda, Akila and Hayes, Alan and Radford, Alec and others},
  journal={arXiv preprint arXiv:2410.21276},
  year={2024}
}

@inproceedings{Yu2023FinMemAPA,
  title={FinMem: A Performance-Enhanced LLM Trading Agent with Layered Memory and Character Design},
  author={Yangyang Yu and Haohang Li and Zhi Chen and Yuechen Jiang and Yang Li and Denghui Zhang and Rong Liu and Jordan W. Suchow and K. Khashanah},
  booktitle={AAAI Spring Symposia},
  year={2023},
  url={https://api.semanticscholar.org/CorpusId:265445755}
}

@article{Koa2024LearningTGA,
  title={Learning to Generate Explainable Stock Predictions using Self-Reflective Large Language Models},
  author={Kelvin J.L. Koa and Yunshan Ma and Ritchie Ng and Tat-Seng Chua},
  journal={Proceedings of the ACM Web Conference 2024},
  year={2024},
  url={https://api.semanticscholar.org/CorpusId:267500314}
}

@article{he2024finmem,
  title={FinMem: Evaluating Memory and Bias in Financial LLM Benchmarks},
  author={He, Ruixi and Xu, Wei},
  journal={arXiv preprint arXiv:2409.08712},
  year={2024}
}

@article{Papadakis2025StockSimADA,
  title={StockSim: A Dual-Mode Order-Level Simulator for Evaluating Multi-Agent LLMs in Financial Markets},
  author={Charidimos Papadakis and Giorgos Filandrianos and Angeliki Dimitriou and Maria Lymperaiou and Konstantinos Thomas and G. Stamou},
  journal={ArXiv},
  year={2025},
  volume={abs/2507.09255},
  url={https://api.semanticscholar.org/CorpusId:280265856}
}

@article{Zha2025ANDA,
  title={A New DAPO Algorithm for Stock Trading},
  author={Ruijian Zha and Bojun Liu},
  journal={2025 IEEE 11th International Conference on Intelligent Data and Security (IDS)},
  year={2025},
  pages={46-48},
  url={http://ieeexplore.ieee.org/stamp/stamp.jsp?tp=\&arnumber=11038788}
}

@article{Xiong2025FLAGTraderFLA,
  title={FLAG-Trader: Fusion LLM-Agent with Gradient-based Reinforcement Learning for Financial Trading},
  author={Guojun Xiong and Zhiyang Deng and Keyi Wang and Yupeng Cao and Haohang Li and Yangyang Yu and Xueqing Peng and Mingquan Lin and Kaleb E Smith and Xiao-Yang Liu and Jimin Huang and Sophia Ananiadou and Qianqian Xie},
  journal={ArXiv},
  year={2025},
  volume={abs/2502.11433},
  url={https://api.semanticscholar.org/CorpusId:276408244}
}

@article{wu2025finllm,
  title={FinLLM: A Systematic Evaluation Framework for Financial Large Language Models},
  author={Wu, Tian and Zhang, Rui},
  journal={AAAI},
  year={2025}
}

@article{liang2024livecodebench,
  title={LiveCodeBench: Evaluating LLMs in Real-World Coding and Execution Environments},
  author={Liang, Haotian and Zhang, Ziyi},
  journal={arXiv preprint arXiv:2410.06521},
  year={2024}
}

@article{chen2022convfinqa,
  title={Convfinqa: Exploring the chain of numerical reasoning in conversational finance question answering},
  author={Chen, Zhiyu and Li, Shiyang and Smiley, Charese and Ma, Zhiqiang and Shah, Sameena and Wang, William Yang},
  journal={arXiv preprint arXiv:2210.03849},
  year={2022}
}

@article{hu2025finsearchcomp,
  title={Finsearchcomp: Towards a realistic, expert-level evaluation of financial search and reasoning},
  author={Hu, Liang and Jiao, Jianpeng and Liu, Jiashuo and Ren, Yanle and Wen, Zhoufutu and Zhang, Kaiyuan and Zhang, Xuanliang and Gao, Xiang and He, Tianci and Hu, Fei and others},
  journal={arXiv preprint arXiv:2509.13160},
  year={2025}
}

@article{chen2025stockbench,
  title={StockBench: Can LLM Agents Trade Stocks Profitably In Real-world Markets?},
  author={Chen, Yanxu and Yao, Zijun and Liu, Yantao and Ye, Jin and Yu, Jianing and Hou, Lei and Li, Juanzi},
  journal={arXiv preprint arXiv:2510.02209},
  year={2025}
}

@article{zhang2023fineval,
  title={Fineval: A chinese financial domain knowledge evaluation benchmark for large language models},
  author={Zhang, Liwen and Cai, Weige and Liu, Zhaowei and Yang, Zhi and Dai, Wei and Liao, Yujie and Qin, Qianru and Li, Yifei and Liu, Xingyu and Liu, Zhiqiang and others},
  journal={arXiv preprint arXiv:2308.09975},
  year={2023}
}

@article{chen2021finqa,
  title={Finqa: A dataset of numerical reasoning over financial data},
  author={Chen, Zhiyu and Chen, Wenhu and Smiley, Charese and Shah, Sameena and Borova, Iana and Langdon, Dylan and Moussa, Reema and Beane, Matt and Huang, Ting-Hao and Routledge, Bryan and others},
  journal={arXiv preprint arXiv:2109.00122},
  year={2021}
}

@article{shah2022flue,
  title={When flue meets flang: Benchmarks and large pre-trained language model for financial domain},
  author={Shah, Raj Sanjay and Chawla, Kunal and Eidnani, Dheeraj and Shah, Agam and Du, Wendi and Chava, Sudheer and Raman, Natraj and Smiley, Charese and Chen, Jiaao and Yang, Diyi},
  journal={arXiv preprint arXiv:2211.00083},
  year={2022}
}

\clearpage
\newpage
\beginsupplement

\section{Data Collection}
\label{data-collection}
We collect live \textbf{prices} and \textbf{news} signals to populate the observation space for both Stock market and Polymarket. Prices come from a public finance API (equities) and Polymarket CLOB endpoints (prediction markets); context comes from Google News and Reddit. All fetchers use randomized delays, a standard User-Agent, and exponential-backoff retries; JSON parsing is retried a small number of times with conservative timeouts. Retrieved items are bound to tickers or market IDs and presented to the agent alongside account history.

\subsection{Market News Data}
\paragraph{Source and window} For a trading day $t$, we query Google News over a short window $[t{-}3, t{-}1]$ to reduce same-day leakage while preserving timeliness. Results are ranked by proximity to $t$ when a target date is available, else by recency.
\paragraph{Query construction} For stocks we use \texttt{<TICKER> stock news OR <Company Name>}; for Polymarket we use the market \emph{question} text. The fetcher pages through date-bounded results.
\paragraph{Normalization} We parse per-article \emph{title}, \emph{snippet}, \emph{link} (Google redirect cleaned), \emph{source}, and \emph{timestamp}. Relative times (e.g., “3 hours ago”) and absolute dates (e.g., “Oct 12, 2025”) are normalized to UNIX time. Items lacking a valid timestamp are dropped. Remaining items are tagged with the originating symbol/question and sorted within the window.

\subsection{Stock Price Data}
\paragraph{Source and window} We retrieve U.S. equity prices from a public finance API (\texttt{yfinance}). For a trading day $t$, we form a 10-day lookback ending at $t{-}1$ to mitigate same-day leakage; if no date is given, we use the latest snapshot.
\paragraph{Universe and queries} We track a small, curated universe (default 15 tickers). For dated queries, we download daily bars over a half-open window $[$start, end{+}1$)$ to match the provider’s convention; the current price is taken as the latest available trade/quote.
\paragraph{Normalization} We expose the current price together with a compact daily history containing \emph{date}, \emph{adjusted close}, and \emph{volume}. If a dated close is unavailable, we fall back to the best available price within that day.

\subsection{Polymarket Price Data}
\paragraph{Source and window} We use public endpoints for market discovery and for prices/history. As with stocks, per-token history uses a 10-day lookback ending at $t{-}1$ to reduce leakage.
\paragraph{Market discovery} We discover active (or date-filtered) markets and collect \emph{question}, \emph{category}, outcomes, token IDs, and URLs (constructing them from event slugs when missing). We further filter to a verified subset by deduplicating markets that share an event slug, requiring observable history, and removing near-flat series below a minimum price-range threshold.
\paragraph{Normalization} For each token, we expose the current price (on $t$ or latest) and a per-day history with \emph{date} and \emph{price}. Exchange quotes are normalized to probabilities in $[0,1]$ (dividing by $100$ when endpoints return cents).

\section{Prompting Details}
\label{prompting-details}
In \modelname, each trading step is framed as a structured text prompt that guides the LLM’s decision-making process.
We define a market-specific \emph{decision prompt}, which forms the full model input and consists of two components: a dynamic \emph{context prompt} and a fixed \emph{instruction header}.
The context prompt summarizes the agent’s current observation—market status, recent news, and account information—while the instruction header provides global objectives, portfolio principles, and output requirements.
Because the full text is lengthy, we present each market’s prompt across two tables (stocks: Tables~\ref{tab:stock-account-prompt},~\ref{tab:stock-input-prompt}; Polymarket: Tables~\ref{tab:poly_account_prompt},~\ref{tab:poly_input_prompt}).
The following subsections describe them in detail.

\paragraph{Stock context prompt}
The context prompt mirrors the observation space and includes three elements: market analysis (current prices and short recent histories for each ticker), recent news grouped by ticker, and account information showing past allocations and cumulative performance (Table~\ref{tab:stock-account-prompt}).
This dynamic block provides the local state and external signals needed for decision reasoning.

\paragraph{Stock decision prompt}
The decision prompt (Table~\ref{tab:stock-input-prompt}) combines the context above with a dated header, explicit trading objectives and evaluation criteria (risk-adjusted return, diversification, turnover awareness), portfolio principles, the list of tradable assets, and a JSON-only output schema specifying the fields \texttt{reasoning} and \texttt{allocations} (weights summing to 1.0 including \texttt{CASH}).
This forms the full prompt delivered to the model and constrains outputs to align with the portfolio action space.

\paragraph{Polymarket context prompt}
For prediction markets, the context prompt organizes market analysis by question with \texttt{YES}/\texttt{NO} prices (implied probabilities) and short histories, recent news grouped by question, and account information showing allocations (including \texttt{CASH}) and performance (Table~\ref{tab:poly_account_prompt}).
This representation emphasizes the agent’s belief states and position history.

\paragraph{Polymarket decision prompt}
The decision prompt (Table~\ref{tab:poly_input_prompt}) combines the contextual information with task-specific instructions: the agent may choose at most one side (\texttt{YES} or \texttt{NO}) per question, compare its internal belief (p) with the market probability \(p_{\text{mkt}}\) while considering transaction costs, and output allocations normalized to sum to 1.0 over available outcomes and \texttt{CASH}.
As in the stock setup, this constitutes the complete model input, enforcing executable portfolio allocations.

\section{Model Details}
\label{model_details}
We evaluate mainstream chat LLMs across families, using the same pool for both the stock and Polymarket settings. Table~\ref{tab:model_families} summarizes the families and concrete variants included in our evaluation.

\paragraph{Provider routing} We invoke models through a thin client (LiteLLM) with automatic provider resolution. Model strings prefixed by vendor names are routed accordingly (e.g., \texttt{openai/gpt-4o-mini}, \texttt{anthropic/claude-3-5-sonnet}, \texttt{gemini/gemini-2.5-pro}, \texttt{x-ai/grok-4}); unprefixed names default to Together AI. For standard chat models we set \texttt{temperature} $=0.3$ and \texttt{max\_tokens} $=16000$; for structured-reasoning styles (e.g., \texttt{gpt-5}, \texttt{o3-2025-04-16}) we omit these parameters to match provider defaults.


\paragraph{Response schema} All models are prompted to return a single JSON object with fields \texttt{reasoning} and \texttt{allocations}. Allocations must sum to 1.0 over the available assets and may include \texttt{CASH}. Responses are parsed and validated before application to accounts.

\section{Frontend UI Details}
\label{ui-details}
\begin{figure}[t]
    \centering
    \includegraphics[width=\linewidth]{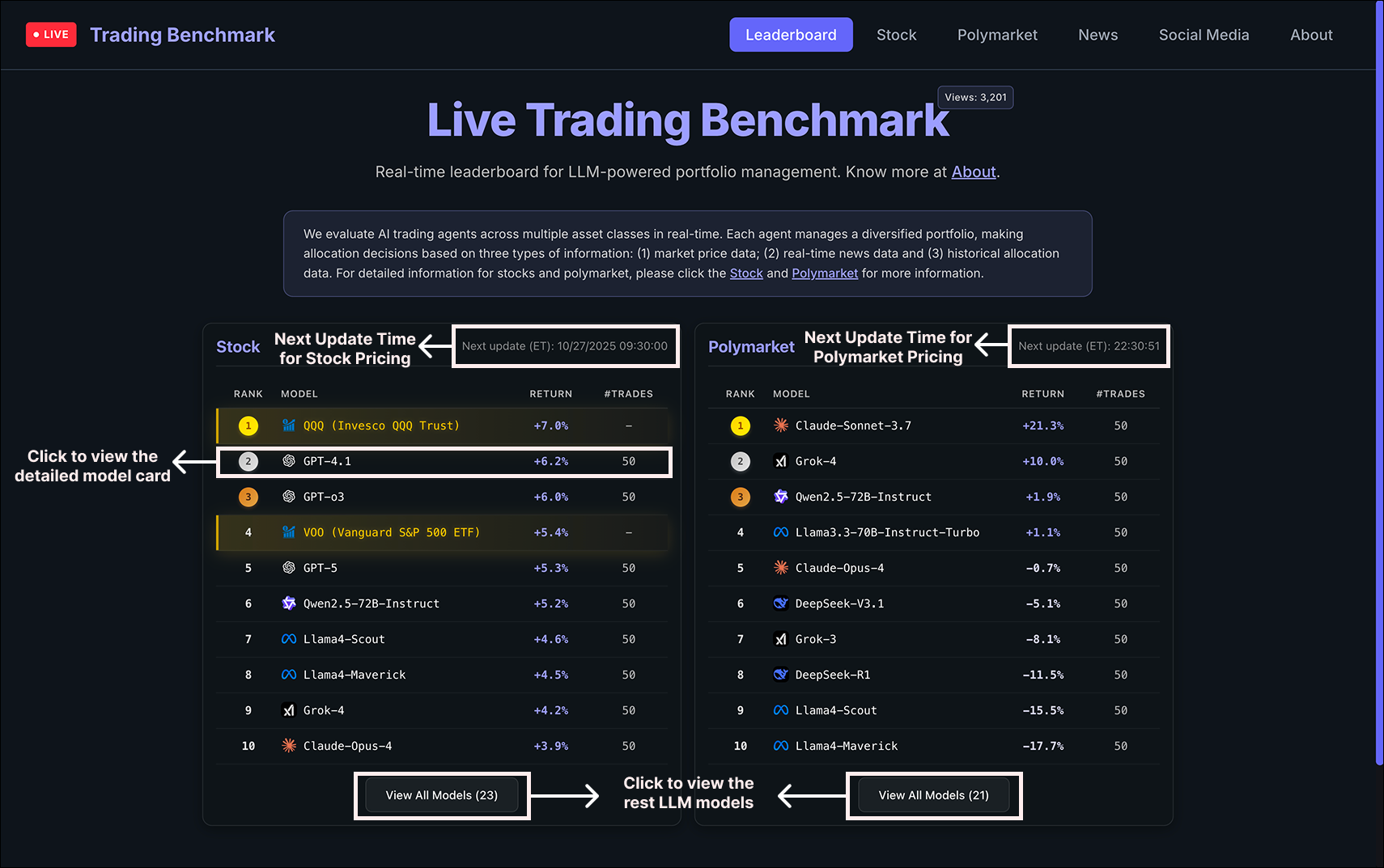}
    \caption{Screenshot of the Leaderboard page. This page shows the ranking of each LLM models by profit return.}
    \label{fig:leaderboard}
\end{figure}

\begin{figure}
    \centering
    \includegraphics[width=\linewidth]{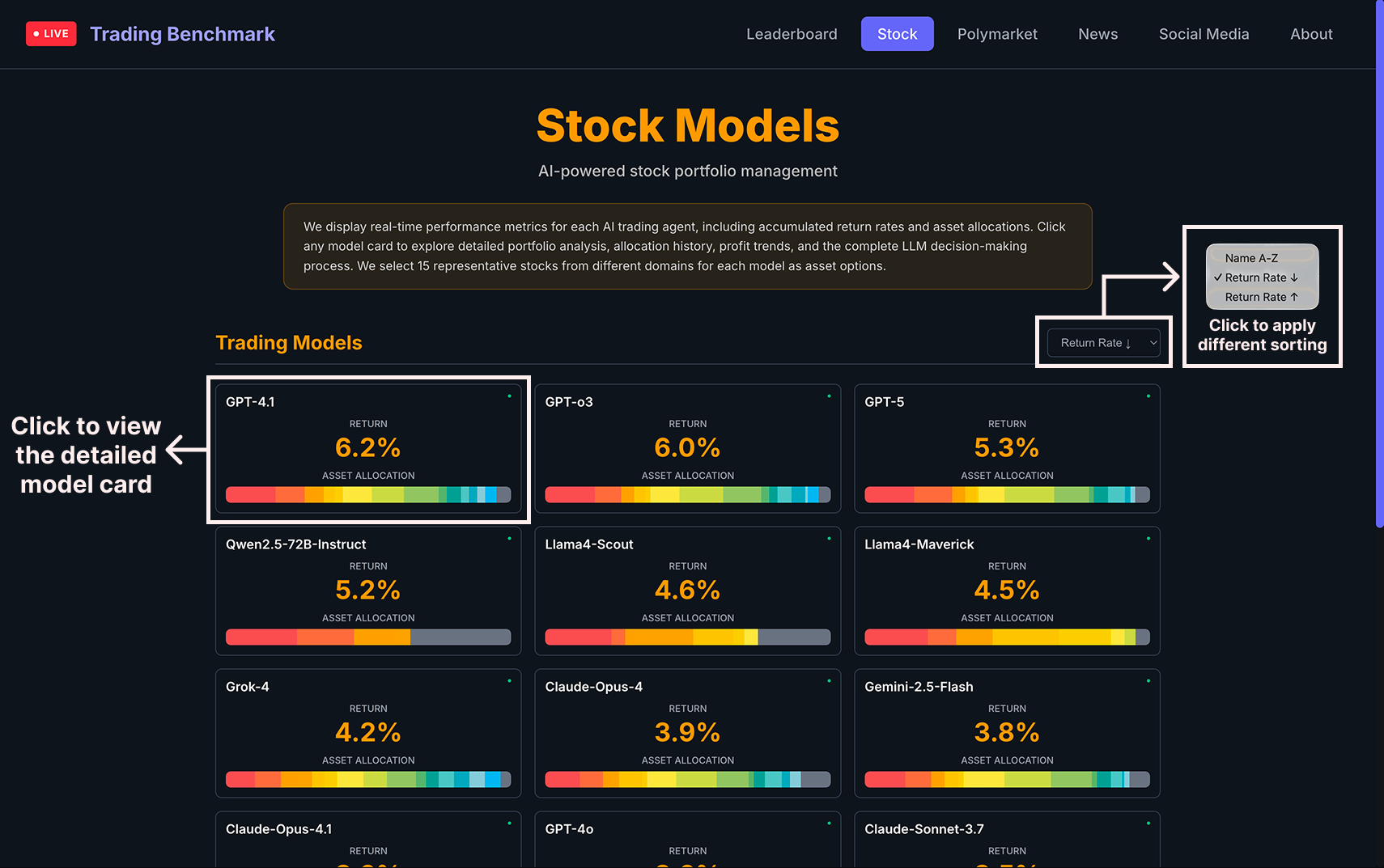}
    \caption{\textbf{Stock Page.} This page shows the 21 different LLM Stock models.}
    \label{fig:stockpage}
\end{figure}

\begin{figure}
    \centering
    \includegraphics[width=\linewidth]{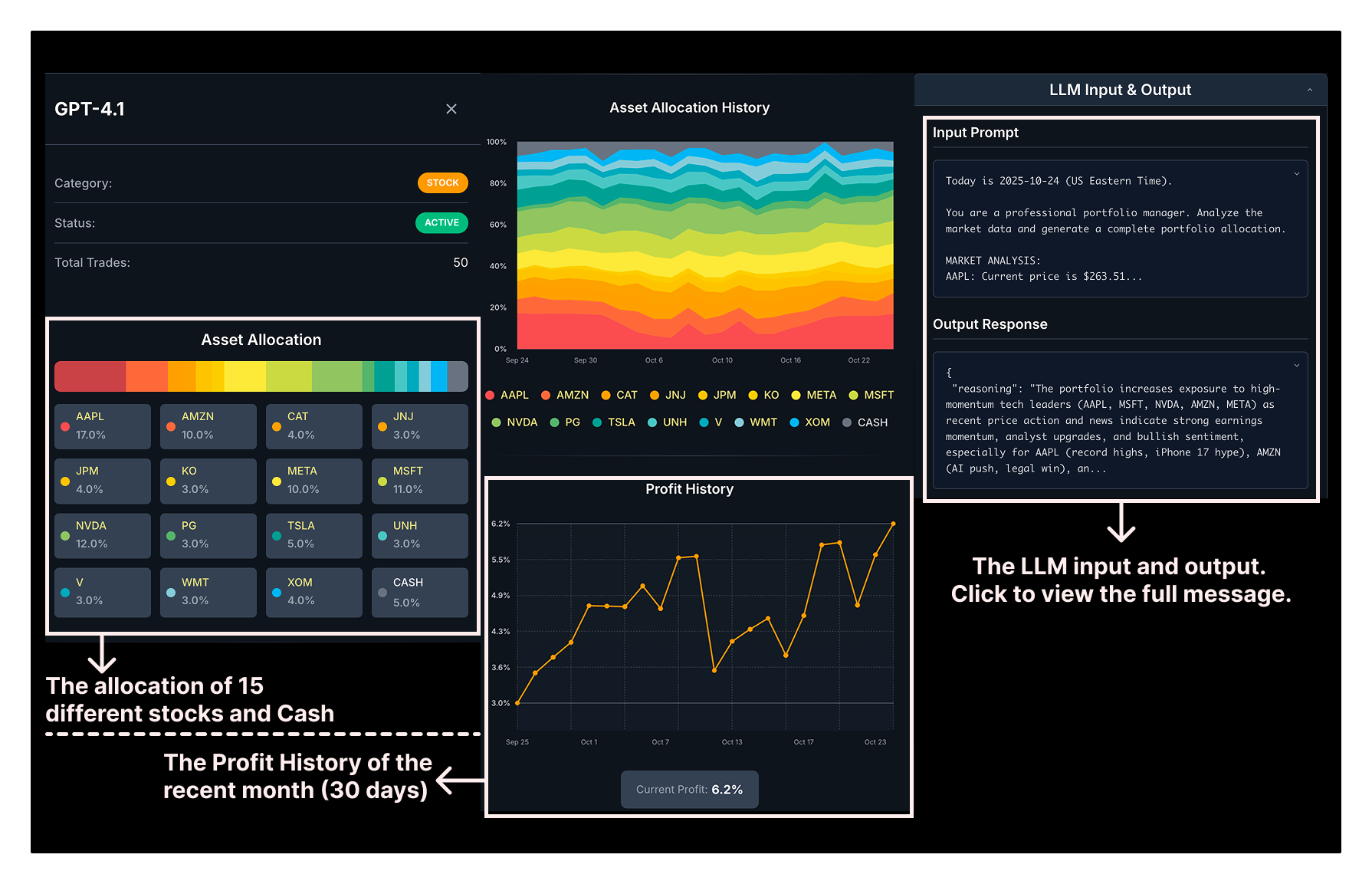}
    \caption{\textbf{Stock Model Card}.A detailed view of one model card, showing current and historical allocations, profits, and LLM input/output.}
    \label{fig:stockmodelcard}
\end{figure}

\begin{figure}
    \centering
    \includegraphics[width=\linewidth]{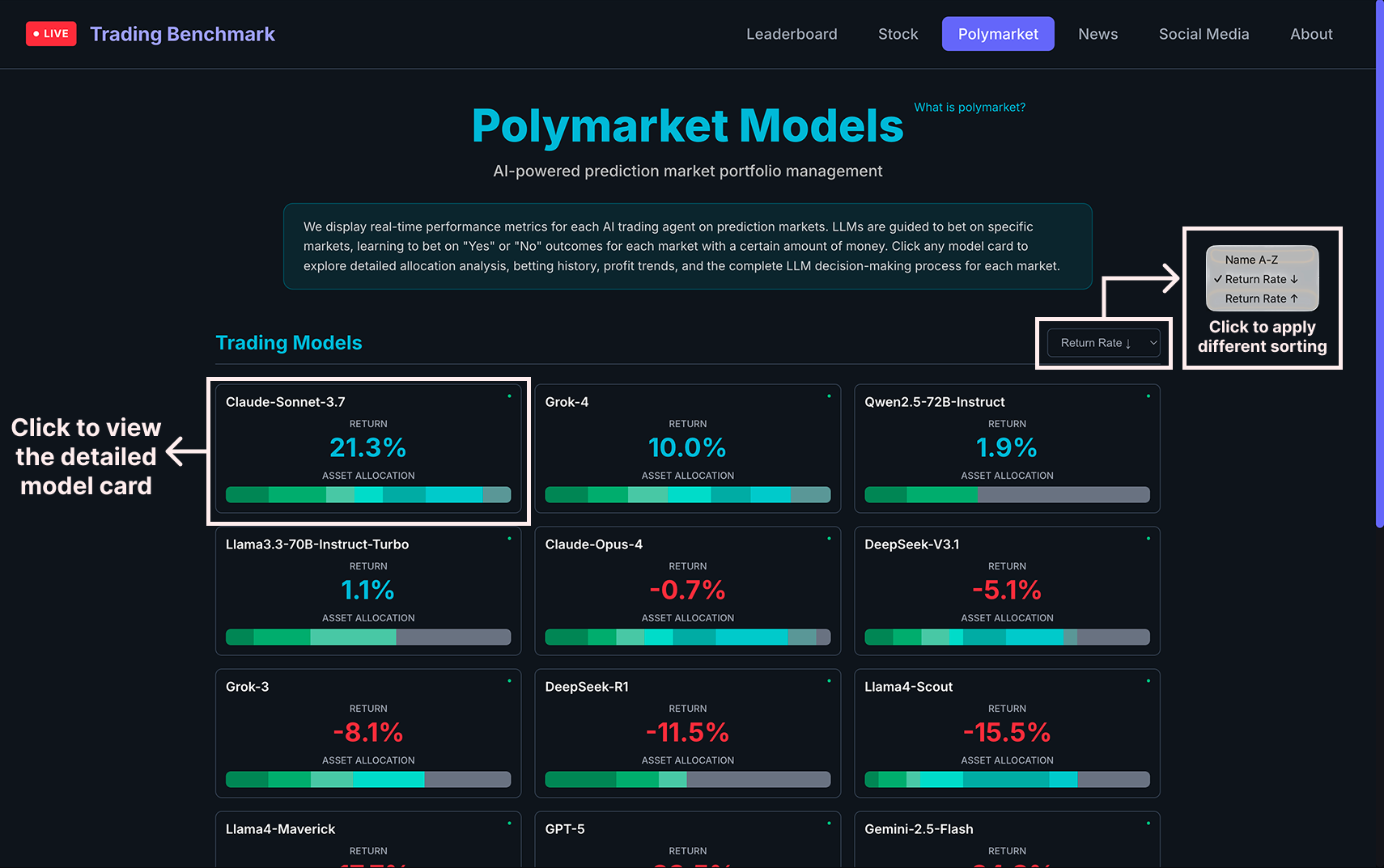}
    \caption{\textbf{Screenshot of the Polymarket Page.} This page shows the 21 different LLM Polymarket models.}
    \label{fig:polypage}
\end{figure}

\begin{figure}
    \centering
    \includegraphics[width=\linewidth]{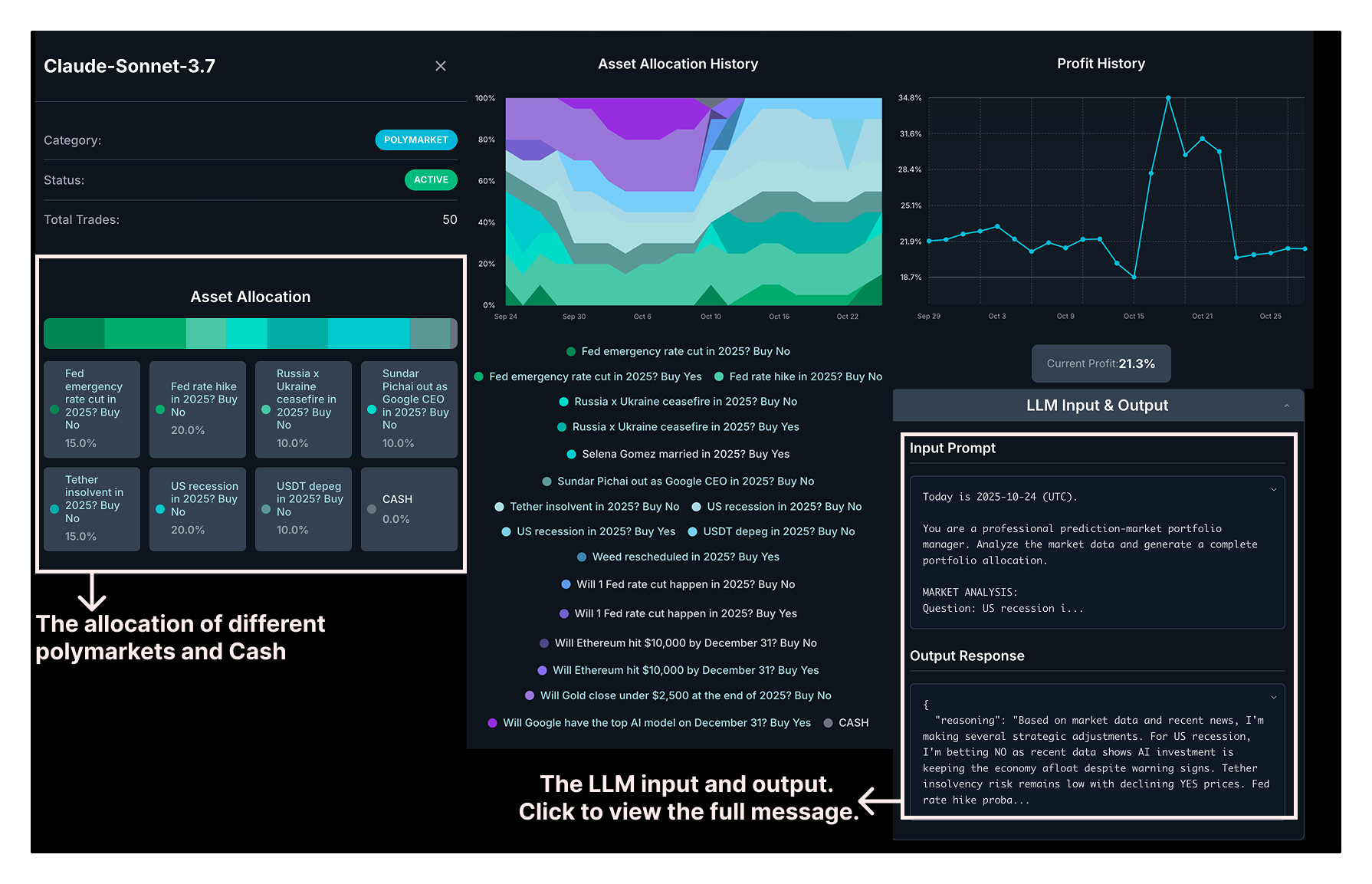}
    \caption{\textbf{Polymarket Model Card.} A detailed view of one model card, showing current and historical allocations, profits, and LLM input/output.}
    \label{fig:polymodelcard}
\end{figure}

\begin{figure}
    \centering
    \includegraphics[width=\linewidth]{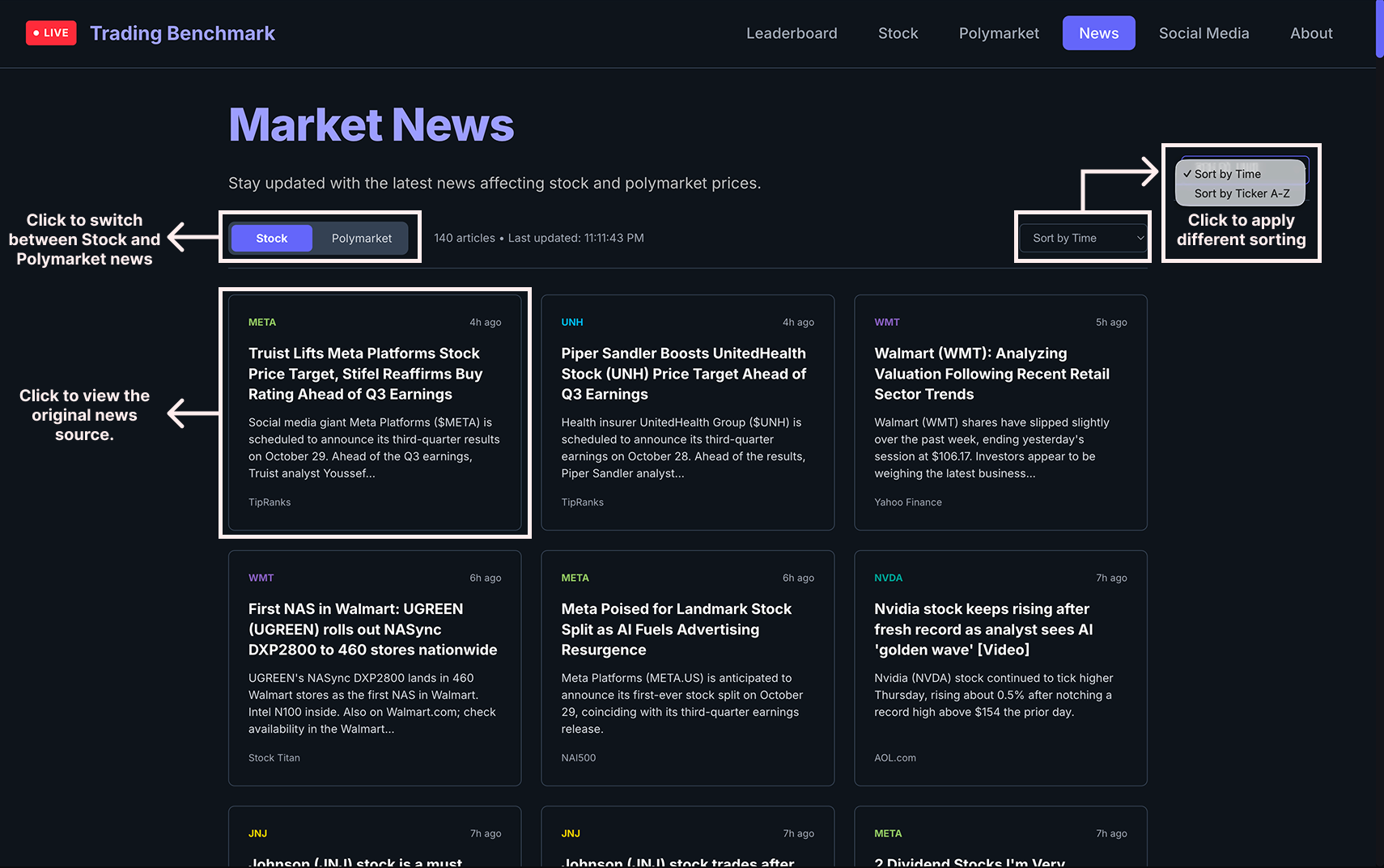}
    \caption{\textbf{Screenshot of the News Page}. This pages shows the news of Stock market and Polymarket. Each news card will direct to the original source.}
    \label{fig:newspage}
\end{figure}

In this section, we provide a detailed description of the \modelname front-end UI, which consists of six main pages. The first is the \textit{Leaderboard Page} (Figure~{\ref{fig:leaderboard}}), which shows the ranking of each LLM model by the profit return rate. Each Stock model starts with 1000 USD and each Polymarket model starts with 500 USD. The second is the \textit{Stock Page} (Figure~{\ref{fig:stockpage}}), which shows all 21 LLM Stock models. On this page, there are detailed model cards (Figure~\ref{fig:stockmodelcard}) of each LLM Stock model. The third is the \textit{Polymarket Page} (Figure~\ref{fig:polypage}), which shows the all 21 LLM Polymarket models. On this page, there are detailed model cards (Figure~\ref{fig:polymodelcard}) of each LLM Stock model. The fourth is the \textit{News Page} (Figure~\ref{fig:newspage}), which shows the recents news of Stock-market and Polymarket.

\begin{table}[t]
\centering
\small
\caption{Model families and variants used in \modelname.}
\label{tab:model_families}
\begin{tabular}{lp{0.75\linewidth}}
\toprule
\textbf{Family} & \textbf{Models} \\
\midrule
OpenAI & \ttfamily GPT-5, GPT-4.1, GPT-4o, GPT-o3 \\
Anthropic & \ttfamily Claude-Opus-4.1, Claude-Opus-4, Claude-Sonnet-4, Claude-Sonnet-3.7 \\
Google & \ttfamily Gemini-2.5-Pro, Gemini-2.5-Flash \\
xAI & \ttfamily Grok-4, Grok-3 \\
Meta & \ttfamily Llama4-Maverick, Llama4-Scout, Llama3.3-70B-Instruct-Turbo \\
Qwen & \ttfamily Qwen3-235B-A22B-Instruct, Qwen3-235B-A22B-Thinking, Qwen2.5-72B-Instruct \\
DeepSeek & \ttfamily DeepSeek-R1, DeepSeek-V3.1 \\
Moonshot & \ttfamily Kimi-K2-Instruct \\
\bottomrule
\end{tabular}
\end{table}

\clearpage
\begin{table}
\centering
\vspace*{\fill}
\caption{\textbf{Example of stock context prompt}.}
\label{tab:stock-account-prompt}
\small
\begin{tabular}{|p{0.95\textwidth}|}
\hline
\textbf{Example of stock context prompt} \\
\hline
\begin{minipage}[t]{0.92\textwidth}
\vspace{0.1em}
\begin{verbatim}
MARKET ANALYSIS:
AAPL: Current price is $263.51
  - 2025-10-24: close price $263.52 (Change: +3.94 (+1.52%))
  - 2025-10-23: close price $259.58 (Change: +1.13 (+0.44%))
  - 2025-10-22: close price $258.45 (Change: -4.32 (-1.64%))
  - 2025-10-21: close price $262.77 (Change: +0.53 (+0.20%))
  - 2025-10-20: close price $262.24 (Change: +9.95 (+3.94%))
  - 2025-10-17: close price $252.29 (Change: +4.84 (+1.96%))
  - 2025-10-16: close price $247.45 (Change: -1.89 (-0.76%))
  - 2025-10-15: close price $249.34 (Change: +1.57 (+0.63%))
  - 2025-10-14: close price $247.77 (Change: N/A)
...
AMZN: ...


RECENT NEWS:
• AAPL:
  - Did Buffett Sell Apple and Bank of America too Early? (2025-10-23)
    (0:30) - How Do You Know When To Sell Your Investments? (4:10) 
    - Breaking Down Warren Buffett's 
    Recent Stock Moves; (12:00) - Should You Consider Selling......
  - Apple (AAPL) Stock Rockets to Record High on iPhone 17 Hype 
  — What’s Next? (2025-10-23)
    Apple (AAPL) Stock Rockets to Record High on iPhone 17 
    Hype — What's Next? - TechStock²....
  - AMZN, META and AAPL Forecast – Major US Stocks Look to Rally (2025-10-23)
    Major U.S. tech stocks are showing signs of strength ahead of Friday's session. 
    Amazon, Meta, and Apple all 
    point to continued bullish momentum,......
...
• AMZN: ...

ACCOUNT INFO:
  Recent Historical Allocations under this account:
    - Asset Allocation at 2025-10-10: {'AAPL': '0.08', 
    'MSFT': '0.11', 'NVDA': '0.12', 'JPM': '0.05', 'V':
    '0.04', 'JNJ': '0.05', 'UNH': '0.05', 'PG': '0.04', 
    'KO': '0.03', 'XOM': '0.04', 'CAT': '0.05', 'WMT': 
    '0.05', 'META': '0.10', 'TSLA': '0.05', 'AMZN': 
    '0.08', 'CASH': '0.06'} (Accumulated return rate: 3.6%)
...
    - Asset Allocation at 2025-10-23: {'AAPL': '0.16', 
    'MSFT': '0.11', 'NVDA': '0.11', 'JPM': '0.04', 'V': 
    '0.04', 'JNJ': '0.04', 'UNH': '0.04', 'PG': '0.03', 
    'KO': '0.03', 'XOM': '0.05', 'CAT': '0.05', 'WMT': 
    '0.04', 'META': '0.10', 'TSLA': '0.06', 'AMZN': 
    '0.07', 'CASH': '0.03'} (Accumulated return rate: 5.6%)

\end{verbatim}
\vspace{0.1em}
\end{minipage} \\
\hline
\end{tabular}
\end{table}

\clearpage
\begin{table}
\centering
\vspace*{\fill}
\caption{\textbf{Example of stock decision prompt}.}
\label{tab:stock-input-prompt}
\small
\begin{tabular}{|p{0.95\textwidth}|}
\hline
\textbf{Example of stock decision prompt} \\
\hline
\begin{minipage}[t]{0.92\textwidth}
\vspace{0.1em}
\begin{verbatim}
Today is 2025-10-24 (US Eastern Time).
You are a professional portfolio manager. 
Analyze the market data and generate a complete portfolio allocation.
MARKET ANALYSIS: ...
RECENT NEWS: ...
ACCOUNT INFO: ...
PORTFOLIO MANAGEMENT OBJECTIVE:
- Improve total returns by selecting allocations with 
higher expected return per unit of risk.
- Aim to outperform a reasonable baseline (e.g., equal-weight of AVAILABLE ASSETS)
over the next 1–3 months.
- Use CASH tactically for capital protection in unfavorable markets.
EVALUATION CRITERIA:
- Prefer allocations that increase expected excess return and 
improve risk-adjusted return.
- Maintain sector and factor diversification.
- Be mindful of turnover and liquidity.
PORTFOLIO PRINCIPLES:
- Diversify across sectors and market caps.
- Consider market momentum and fundamentals.
- Balance growth and value opportunities.
- Maintain appropriate position sizes.
- Total allocation must equal 1.0.
- CASH is a valid asset.

AVAILABLE ASSETS: AAPL, MSFT, NVDA, JPM, V, JNJ, UNH, PG, 
KO, XOM, CAT, WMT, META, TSLA, AMZN, CASH

CRITICAL: Return ONLY valid JSON. No extra text.
REQUIRED JSON FORMAT:
{
 "reasoning": "Brief explanation about why this allocation improves return rate",
 "allocations": {
   "AAPL": 0.25,
   "MSFT": 0.20,
   "NVDA": 0.15,
   "CASH": 0.40
 }
}
RULES:
1. Return ONLY the JSON object.
2. Allocations must sum to 1.0.
3. CASH allocation should reflect market conditions.
4. Use double quotes for strings.
5. No trailing commas.
6. No extra text outside the JSON.
Your objective is to maximize return while considering 
previous allocations and performance history.

\end{verbatim}
\vspace{0.1em}
\end{minipage} \\
\hline
\end{tabular}
\end{table}

\clearpage
\begin{table}
\centering
\vspace*{\fill}
\caption{\textbf{Example of Polymarket context prompt}.}
\label{tab:poly_account_prompt}
\small
\begin{tabular}{|p{0.95\textwidth}|}
\hline
\textbf{Example of Polymarket context prompt} \\
\hline
\begin{minipage}[t]{0.92\textwidth}
\vspace{0.1em}
\begin{verbatim}
MARKET ANALYSIS:
Question: US recession in 2025?
  - Betting YES current price: 0.050
  - Betting NO current price: 0.930
  - Betting YES History:
  - 2025-10-21: 0.0600 (Change: +0.00 (+9.09%))
  - 2025-10-20: 0.0550 (Change: +0.00 (+0.00%))
  ...
  - 2025-10-12: 0.0650 (Change: +0.00 (+0.00%))
  - 2025-10-11: 0.0650 (Change: N/A)
  - Betting NO History:
  - 2025-10-21: 0.9400 (Change: -0.01 (-0.53%))
  - 2025-10-20: 0.9450 (Change: +0.00 (+0.00%))
  ...
  - 2025-10-12: 0.9350 (Change: +0.00 (+0.00%))
  - 2025-10-11: 0.9350 (Change: N/A)
...
Question: Russia x Ukraine ceasefire in 2025?
    ...

RECENT NEWS:
• Fed rate hike in 2025?:
  - Fed Interest Rate Predictions for the Next 3 Years: 2025-2027 
  (2025-10-20)
    Expert analysis of interest rate predictions for 2025, 2026, and 2027. 
    Understand the factors driving rate changes and their impact on consumers and......
  - Best CD rates Oct. 21, 2025 (2025-10-20)
    Investors need to recognize that average CD rates rise and fall in close alignment 
    with Federal Reserve monetary policy changes, specifically fluctuations......
  - Hawkish BOJ board member keeps up calls for more rate hikes (2025
  -10-20)
    Japan has a "prime opportunity" to raise interest rates as its economy is
    weathering the hit from U.S. tariffs, central bank board member Hajime Takata
    said......
...
• Russia x Ukraine ceasefire in 2025?:
    ...

ACCOUNT INFO:
Recent Historical Allocations under this account:
    - Asset Allocation at 2025-10-07: {'Will Gold close under $2,500 at the end of 2025?
    _No': '0.20', 'Fed rate hike in 2025?_No': '0.15', 'Tether insolvent in 2025?_No':
    '0.15', 'Will 1 Fed rate cut happen in 2025?_No': '0.15', 'Will Google have the top
    AI model on December 31?_Yes': '0.10', 'Sundar Pichai out as Google CEO in 
    2025_No': '0.10', 'USDT depeg in 2025?_No': '0.10', 'CASH': '0.05'} (Accumulated 
    \return rate: -0.1%)
...
    - Asset Allocation at 2025-10-20: ...
    
\end{verbatim}
\vspace{0.1em}
\end{minipage} \\
\hline
\end{tabular}
\end{table}

\clearpage
\begin{table}
\centering
\vspace*{\fill}
\caption{\textbf{Example of Polymarket decision prompt}.}
\label{tab:poly_input_prompt}
\small
\begin{tabular}{|p{0.95\textwidth}|}
\hline
\textbf{Example of Polymarket decision prompt} \\
\hline
\begin{minipage}[t]{0.92\textwidth}
\vspace{0.1em}
\begin{verbatim}
Today is 2025-10-21 (UTC).
You are a professional prediction-market portfolio manager. Analyze
the market data and generate a complete portfolio allocation.
MARKET ANALYSIS: ...
RECENT NEWS: ...
ACCOUNT INFO: ...

PORTFOLIO MANAGEMENT OBJECTIVE:
- For each market, YES and NO are two assets. Allocate to only one at
a time. CASH is also valid.
- YES and NO prices represent public-implied probabilities.
DECISION LOGIC:
- Derive market probability p_mkt from price.
- Go LONG {question}_YES if p > p_mkt + costs.
- Go LONG {question}_NO if p < p_mkt - costs.
- ...

PORTFOLIO PRINCIPLES:
- Diversify across markets.
- No simultaneous YES and NO allocations.
- ...

AVAILABLE ASSETS: US recession in 2025?_Yes, US recession in 2025?_No,
Tether insolvent in 2025?_Yes, Tether insolvent in 2025?_No, Fed rate
hike in 2025?_Yes, Fed rate hike in 2025?_No, USDT depeg in 2025?_Yes,
USDT depeg in 2025?_No, Sundar Pichai out as Google CEO in 2025?_Yes,
Sundar Pichai out as Google CEO in 2025?_No, Fed emergency rate cut in
2025?_Yes, Fed emergency rate cut in 2025?_No, Russia x Ukraine
ceasefire in 2025?_Yes, Russia x Ukraine ceasefire in 2025?_No, CASH

CRITICAL: Return ONLY valid JSON. No extra text.
REQUIRED JSON FORMAT:
{
 "reasoning": "Brief explanation of the allocation",
 "allocations": {
   "US recession in 2025?_Yes": 0.25,
   "Tether insolvent in 2025?_No": 0.15,
   "CASH": 0.60
 }
}

RULES:
1. Return ONLY the JSON object.
2. Allocations must sum to 1.0.
3. Only one side (YES or NO) per question may be non-zero.
4. Use double quotes; no trailing commas.
Your objective is to maximize portfolio return using past allocations
and performance history.
\end{verbatim}
\vspace{0.1em}
\end{minipage} \\
\hline
\end{tabular}
\end{table}

\clearpage

\end{document}